\documentclass[twocolumn]{aastex631}

\usepackage{graphicx}
\usepackage{natbib}
\usepackage{breakurl}
\usepackage{float}
\usepackage{amsmath}
\usepackage{threeparttable}
\usepackage{placeins}
\usepackage{hyperref}
\usepackage{color}
\usepackage{url}
\usepackage{CJK}

\newcommand{\hst}{\textit{HST}}
\newcommand{\hdha}{HDH$\alpha$}
\newcommand{\lya}{Ly$\alpha$}
\newcommand{\ha}{H$\alpha$}
\newcommand{\hb}{H$\beta$}
\newcommand{\oiii}{[O\,{\sc iii}]}
\newcommand{\oii}{[O\,{\sc ii}]}
\newcommand{\heii}{He\,{\sc ii}}

\shorttitle{The Hubble Deep Hydrogen Alpha Project (HDH$\alpha$)}
\shortauthors{Zhu, Zheng, Rhoads et al.}

\graphicspath{{./}{figures/}}

\begin{document}

\title{The Hubble Deep Hydrogen Alpha (HDH$\alpha$) Project: I. Catalog of Emission-line Galaxies}

\author[0000-0002-2528-0761]{Shuairu Zhu}
\affiliation{Key Laboratory for Research in Galaxies and Cosmology, Shanghai Astronomical Observatory, Chinese Academy of Sciences, 80 Nandan Road, Shanghai 200030, People’s Republic of China}
\affiliation{School of Astronomy and Space Sciences, University of Chinese Academy of Sciences, No. 19A Yuquan Road, Beijing 100049, People’s Republic of China}

\author[0000-0002-9634-2923]{Zhen-Ya Zheng*}
\affiliation{Key Laboratory for Research in Galaxies and Cosmology, Shanghai Astronomical Observatory, Chinese Academy of Sciences, 80 Nandan Road, Shanghai 200030, People’s Republic of China}
\correspondingauthor{Zhen-Ya Zheng}
\email{Email: zhengzy@shao.ac.cn}

\author[0000-0002-1501-454X]{James Rhoads}
\affiliation{Astrophysics Science Division, NASA Goddard Space Flight Center, 8800 Greenbelt Road, Greenbelt, Maryland, 20771, USA}

\author[0000-0002-4419-6434]{Junxian Wang}
\affiliation{CAS Key Laboratory for Research in Galaxies and Cosmology, Department of Astronomy, University of Science and Technology of China, Hefei, Anhui 230026, People’s Republic of China}
\affiliation{School of Astronomy and Space Science, University of Science and Technology of China, Hefei 230026, People’s Republic of China}

\author[0000-0003-4176-6486]{Linhua Jiang}
\affiliation{Kavli Institute for Astronomy and Astrophysics, Peking University, Beijing 100871, People’s Republic of China}


\author[0000-0002-0003-8557]{Chunyan Jiang}
\affiliation{Key Laboratory for Research in Galaxies and Cosmology, Shanghai Astronomical Observatory, Chinese Academy of Sciences, 80 Nandan Road, Shanghai 200030, People’s Republic of China}

\author[0000-0001-6763-5869]{Fang-Ting Yuan}
\affiliation{Key Laboratory for Research in Galaxies and Cosmology, Shanghai Astronomical Observatory, Chinese Academy of Sciences, 80 Nandan Road, Shanghai 200030, People’s Republic of China}

\author[0000-0002-5864-7195]{P. T. Rahna}
\affiliation{Key Laboratory for Research in Galaxies and Cosmology, Shanghai Astronomical Observatory, Chinese Academy of Sciences, 80 Nandan Road, Shanghai 200030, People’s Republic of China}

\author[0000-0003-3424-3230]{Weida Hu}
\affiliation{CAS Key Laboratory for Research in Galaxies and Cosmology, Department of Astronomy, University of Science and Technology of China, Hefei, Anhui 230026, People’s Republic of China}

\author[0000-0003-3987-0858]{Ruqiu Lin}
\affiliation{Key Laboratory for Research in Galaxies and Cosmology, Shanghai Astronomical Observatory, Chinese Academy of Sciences, 80 Nandan Road, Shanghai 200030, People’s Republic of China}
\affiliation{School of Astronomy and Space Sciences, University of Chinese Academy of Sciences, No. 19A Yuquan Road, Beijing 100049, People’s Republic of China}

\author[0000-0001-8534-837X]{Huanyuan Shan}
\affiliation{Key Laboratory for Research in Galaxies and Cosmology, Shanghai Astronomical Observatory, Chinese Academy of Sciences, 80 Nandan Road, Shanghai 200030, People’s Republic of China}

\author{Chun Xu}
\affiliation{Key Laboratory for Research in Galaxies and Cosmology, Shanghai Astronomical Observatory, Chinese Academy of Sciences, 80 Nandan Road, Shanghai 200030, People’s Republic of China}

\author[0000-0001-8581-932X]{Leopoldo Infante}
\affiliation{Las Campanas Observatory, Carnegie Institution of Washington, Casilla 601, La Serena, Chile}

\author[0000-0003-0151-0718]{L. Felipe Barrientos}
\affiliation{Instituto de Astrofísica and Centro de Astroingeniería, Facultad de Física, Pontificia Universidad Católica de Chile, Casilla 306, Santiago 22, Chile}

\author[0000-0003-3728-9912]{Xianzhong Zheng}
\affiliation{Purple Mountain Observatory, Chinese Academy of Sciences, 10 Yuanhua Road, Nanjing 210023, People’s Republic of China}

\author[0000-0001-9694-2171]{Guanwen Fang}
\affiliation{School of Mathematics and Physics, Anqing Normal University, Anqing 246011, People’s Republic of China}

\author[0000-0002-2384-3436]{Zhixiong Liang}
\affiliation{School of Mathematics and Physics, Anqing Normal University, Anqing 246011, People’s Republic of China}


\begin{abstract}

We present the first results of the Hubble Deep Hydrogen Alpha (HDH$\alpha$) project, 
which analyzes the space-borne deep H$\alpha$ narrowband imaging data in the GOODS-S region. 
The HDH$\alpha$ data comprises 72 orbits' images taken with the \textit{HST} ACS/WFC F658N filter.
The exposure time varies across a total area of $\sim$76.1 $\rm{arcmin}^2$, adding up to 
a total exposure time of 195.7 ks, among which 68.8 ks are spent in the deepest region. These images are aligned, 
reprojected, and combined to have the same pixel grid as the Hubble Legacy Fields (HLF). 
The scientific goals of the \hdha\ include establishing a sample of emission-line galaxies (ELGs) including \oiii\ emitters at $z\sim$ 0.3, \oii\ emitters at $z\sim$ 0.8, and Lyman-$\alpha$ emitters (LAEs) at $z \sim 4.4$, studying the line morphology of ELGs with high resolution imaging data, and statistically analyzing the line luminosity functions and line equivalent-width distributions of ELGs selected with \hst. 
Furthermore, the HDH$\alpha$ project enhances the legacy value of the GOODS-S field by contributing the first \hst-based narrowband image to the existing data sets, which includes the \hst\ broadband data and 
other ancillary data from X-ray to radio taken by other facilities. In this paper, we describe the data reduction process of the \hdha, select ELGs based on \hst's F658N and broadband data, validate the redshifts of the selected candidates by cross matching with the public spectroscopic catalogs in the GOODS-S, and present a final catalog of the confirmed \oiii\ emitters at $z\sim$ 0.3, \oii\ emitters at $z\sim$ 0.8, and LAEs at $z \sim 4.4$.

\end{abstract}

\section{Introduction} 
\label{sec:intro}

Distant galaxies are essential for our understanding of the early Universe and the formation and evolution of galaxies.
In the optical band, there are two techniques widely used to search for and study high redshift galaxies, which are the Lyman break technique and the emission line excess technique. The former technique, also called the drop-out 
technique \citep[][]{Steidel1996}, selects high-$z$ galaxies via their rest-frame UV break features in series broadband images, caused by the absorption of neutral hydrogen within the galaxies and in the intergalactic medium 
(IGM). The latter technique identifies galaxies with a strong emission line at a particular redshift, where the central wavelength of the emission line is redshifted in the bandpass of the narrowband filter, causing a flux excess in the narrowband over the corresponding broadband. This
method identifies high redshift galaxies based on their strong Lyman-$\alpha$ emission line, which was predicted 
as a promising tracer of high redshift galaxies in the 1960s \citep{Partridge1967}. However, this technique did not succeed until the middle 1990s when better CCD detectors with larger field-of-view (FOV), larger telescopes, and better narrowband filters were available \citep{Cowie1998, Rhoads2000}.

Over the past two decades, the drop-out technique has been widely used to select high redshift star-forming galaxies 
with deep broadband images from numerous survey campaigns, especially with the Hubble Space Telescope (\textit{HST}). 
With the images taken in the first deep blank field, the Hubble Deep Field \citep{Williams1996}, galaxies at 
$z \sim 4$ were selected successfully to study cosmic star formation history \citep{Madau1996}. Using the deep mosaics based on images delivered by  
the Great Observatories Origins Deep Survey \citep[GOODS, ][]{Giavalisco04} and Hubble Ultra Deep Field program 
\citep[HUDF,][]{Beckwith2006}, the redshift limit had been pushed to $z \sim 6$ \citep{Bouwens2004, Dickinson2004, Bouwens2006}. 
Even a handful of galaxies at $z \sim 7$ have been identified with additional \hst/NICMOS and ground-based (e.g., $VLT$/ISAAC, 
$Subaru$/MOIRCS) observations \citep{Bouwens2008}. Further breakthroughs in the search of high-$z$ galaxies were accomplished using the Wide Field Camera 3 Infrared Channel (WFC3/IR). 
New generation surveys taken with the WFC3/IR, including the HUDF09 \citep{Bouwens2011}, the Cosmic Assembly Near-infrared Deep Extragalactic Legacy Survey \citep[CANDELS, ][]{Grogin2011, Koekemoer2011}, the Early Release Science program of WFC3 \citep[ERS,][]{Windhorst2011},
and the HUDF12 \citep{Koekemoer2013}, had enabled the discovery of galaxies 
up to $z\simeq 8-10$ \citep{Bouwens2010,Oesch2013, Oesch2014, Bouwens2015b}. Recently, the \textit{HST} surveys of gravitational lensing galaxy clusters have provided a new way to search for high $z$ galaxies \citep[e.g. the Hubble Frontier Field,][]{Lotz2017}, which used galaxy clusters as a natural telescope to search and study high-$z$ galaxies \citep[][]{McLeod2016, Ishigaki2018, Bouwens2022}.

Unlike the tremendously successful broadband drop-out surveys with \textit{HST} as mentioned above, 
emission-line surveys using narrowband selection technique are mainly carried out with ground-based telescopes, 
e.g., the Subaru/XMM-Newton Deep Survey \citep[SXDS,][]{ouchi2008}, the Lyman Alpha Galaxies in the Epoch of Reionization Survey \citep[LAGER,][]{zheng2017}, 
the Subaru SILVERRUSH survey \citep[][]{ouchi2018}, and the upcoming Javalambre-Physics of the Accelerated Universe Astrophysical Survey \citep[J-PAS,][]{Benitez2014}. 
A wide range of research topics on Lyman-$\alpha$ Emitters (LAEs) had been probed based on the narrowband imaging technique, including diffuse Lyman-$\alpha$ halos \citep[e.g.][]{Momose2014, Momose2016}, 
giant diffuse Lyman-$\alpha$ nebulae \citep[][]{Cantalupo2014, Hennawi2015, Cai2017, Battaia2018}, overdense regions 
such as protoclusters and large scale structures at high redshifts \citep{zheng2016, Cai2017a, Jiang2018, Hu2020,xzzheng2021}, and the ionization state of the cosmic reionization epoch \citep[][]{Malhotra2002, ouchi2010, kashikawa2011, zheng2017, Hu2019, Wold2021}.  Moreover, this technique has also been used to search for other emission-line emitters to constrain the line luminosity functions and the cosmic star formation rate densities at $z\lesssim2$.\citep[][]{Sobral13, Hayashi2018, Khostovan2020}.

With \hst, narrowband imaging surveys for high-$z$ galaxies are rare.
Most of \textit{HST}'s narrowband imaging surveys were targeted at local galaxies and objects, e.g., local ultra-faint dwarf galaxies \citep{Fu2021}, young stellar objects \citep{Ferreras2009}, and supernova remnants \citep[i.e., SN1987A,][]{Larsson2013} in the Large Magellanic Cloud.
In the high-$z$ Universe, only a few of \hst's narrowband surveys were reported. In the \lya\ morphological studies of 
high-$z$ LAEs, \citet{Bond2010} reported compact \lya\ emission of LAEs at $z\sim$ 3.1 with the WFPC2 narrowband observations, while \citet{Finkelstein2011} detected the resolved \lya\ emission from two out of three LAEs at $z \sim 4.4$ with the ACS/WFC narrowband observations. \citet{Decarli2012} used the WFC3/UVIS narrowband observations to search for the extended \lya\ emission around two quasars at $z>6$, while neither extended \lya\ emission around quasars nor LAE companions were found. 

Here we introduce the Hubble Deep Hydrogen Alpha (\hdha) project, which is based on the deepest HST/ACS H$\alpha$ narrowband (F658N) imaging survey data and is also the largest and deepest narrowband imaging survey in the blank field taken with \textit{HST}. 
The final \hdha\ image, covering part of the GOODS-S Field, allows a systematic search for sources with strong emission lines along with deep broadband images at \hst's spatial resolution. 
With this sample, we can further analyze the emission-line galaxies' morphology and spectral energy distribution in the GOODS-S field with the most up-to-date Hubble Legacy Field \citep[HLF: ][Illingworth et al. in prep.]{Illingworth2016} data.

The structure of the paper is as follows. In Section \ref{sec:data}, we describe the observations of the narrowband imaging surveys with \textit{HST} (\S\,\ref{sec:data:hstnb}), the HDH$\alpha$ project (\S\,\ref{sec:data:hdha}), the reduction process of the HDH$\alpha$ narrowband data (\S\,\ref{sec:data:reduc}), and the data quality of the mosaic HDH$\alpha$ image (\S\,\ref{sec:data:quality}).
In Section \ref{sec:photo}, we discuss the photometry of HDH$\alpha$ sources. We select emission-line galaxies (ELGs), including \oiii\ emitters at $z \sim 0.3$, \oii\ emitters at $z \sim 0.8$, and LAEs at $z\sim4.4$ with the HDH$\alpha$ narrowband image and the HLF broadband images in Section \ref{sec:HDHaELGs}.
Finally, we summarize our main results in Section \ref{sec:sum}. Throughout this paper, we adopt a standard cosmology model with parameters 
$\Omega_{M} = 0.3$, $\Omega_{\Lambda} = 0.7$, $H_{0} = \rm{70\ km\ s^{-1} Mpc^{-1}}$.
All magnitudes throughout this paper are given in the AB system \citep{oke1983}.

\section{HST Narrowband Surveys and Data Reduction} 
\label{sec:data}

\subsection{Deep Narrowband Imaging Surveys with HST}
\label{sec:data:hstnb}

We search for the \hst's narrowband imaging surveys in extragalactic fields with the planned and archived exposures catalog (PAEC, version: 2021-12-01)\footnote{\href{https://archive.stsci.edu/hst/paec.html}{https://archive.stsci.edu/hst/paec.html}}, which summarizes all exposures taken or planned with \hst\ since Cycle 7.
We use HEALPix\footnote{\href{http://healpix.sf.net}{http://healpix.sf.net}} \citep[Hierarchical Equal Area isoLatitude Pixelation, 
its python wrapper: \textit{healpy}, see][]{2005ApJ...622..759G, Zonca2019} to divide the celestial sphere into a series of cells with equal areas and calculate the sum of the exposure time of narrowband observations
in each cell. The \textit{healpy}'s parameter $N_\textrm{Side}$ is set to $N_\textrm{Side}$ = 128, corresponding to an area of each cell of 755 $\rm arcmin^2$.

We check the narrowband imaging surveys taken with \hst 's WFPC2, ACS/WFC, WFC3/UVIS, and WFC3/IR using HEALPix. The deepest exposures of a single narrowband filter taken with these cameras are summarized in Tab.~\ref{tab:statsdeepfield}. 
We find that the ACS/WFC has the deepest narrowband (F658N) exposure in GOODS-S, which is a famous extragalactic blank field.
Other deep narrowband surveys in Tab.~\ref{tab:statsdeepfield} are targeted at nearby objects, thus unsuitable for the extragalactic study. Therefore, we do not carry out further reductions to these data sets in this study.

\begin{deluxetable*}{lllll}[tb] 
\label{tab:statsdeepfield}
\tablecaption{The deepest narrowband imaging surveys for each of the \hst\ imaging cameras} \label{tab:stats}
\tablecolumns{5}
\tablehead{\colhead{Camera} & \colhead{Deepest Region} & \colhead{Narrowband Filters} & \colhead{Program Exposure Time} \\
\colhead{} & \colhead{} & \colhead{} & \colhead{(Second)}}
\startdata
\decimalcolnumbers
\textbf{ACS/WFC} & \textbf{R.A}.\boldmath$=03^{h}32^{m}20.62^{s}$\unboldmath, \textbf{Dec.}\boldmath$=-27^{d}57^{m}11.47^{s}$\unboldmath & \colhead{\textbf{F658N}} & \colhead{\textbf{197122}} &\colhead{\textbf{}}\\
{} & \colhead{\textbf{GOODS-S}} & {} & {}\\ 
\hline
WFC3/UVIS & R.A.$=00^{h}25^{m}42.86^{s}$, Dec.$=-72^{d}01^{m}03.45^{s}$ & \colhead{F502N} & \colhead{138186}\\
{} & \colhead{47 Tuc} & \colhead{} & \colhead{}\\
\hline
{WFPC2} & {R.A.$=05^{h}37^{m}30^{s}$, Dec.$=-69^{d}25^{m}21.16^{s}$} & \colhead{F656N} & \colhead{125990} & \colhead{}\\
{} & \colhead{LMC} & \colhead{} & \colhead{}\\
\hline
WFC3/IR & R.A. $=10^{h}38^{m}22.33^{s}$, Dec.$=-58^{d}09^{m}37.33^{s}$ & \colhead{F167N} & \colhead{66291}\\
{} & \colhead{G286.21+0.17} & \colhead{} & \colhead{}\\
\enddata
\end{deluxetable*}

\subsection{HDH$\alpha$: The HST ACS/WFC F658N Imaging Survey in GOODS-S}
\label{sec:data:hdha}

\begin{figure}[h]
    \centering
    \includegraphics[width=0.43\textwidth]{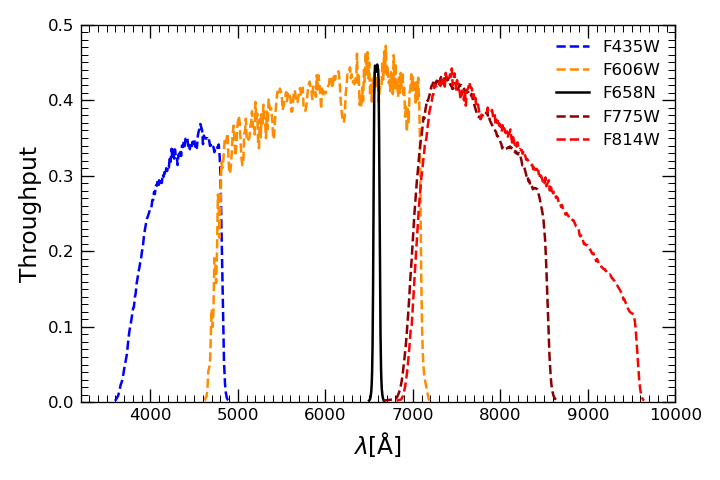}
    \caption{System throughputs of the broadband filters F435W, F606W, F775W, F814W, and the narrowband filter F658N of \textit{HST} ACS/WFC.}
    \label{fig:throughput}
\end{figure}

The F658N narrowband filter is one of the 16 filters onboard the Wide Field Channel (WFC) of \textit{HST}/ACS\footnote{\href{https://etc.stsci.edu/etcstatic/users_guide/appendix_b_acs.html\#wfc-filters}{https://etc.stsci.edu/etcstatic/users\_guide/appendix\_b\_acs.html/ \#wfc-filters}}. 
The F658N filter is described as a H$\alpha$ filter\footnote{See Section 5.1 of ACS Instrument Handbook for Cycle 30, \href{https://hst-docs.stsci.edu/acsihb}{https://hst-docs.stsci.edu/acsihb}} with a central wavelength of 6584\AA\ and a full 
width at half maximum (FWHM) of 87\AA  \footnote{Calculated from \href{https://github.com/spacetelescope/stsynphot_refactor}{stsynphot} \citep{ascl2010}.}, respectively. The system throughput of \hst ACS/WFC F658N, along with broadband filters F435W, F606W, F775W and F814W, is presented in Fig.~\ref{fig:throughput}.

The \hdha\ is located in the GOODS-S field. The narrowband images were collected as parallel observations during the ERS program \citep[Program ID:
11359, PI: Robert O'Connell, ][]{Windhorst2011}. The F658N imaging of the \hdha\ is composed of 8 independent pointings, each of which includes a 9-orbit ACS/WFC F658N observation. It covers a total area of 86 arcmin$^{2}$ in the GOODS-S region. The total exposure time of the \hdha\ is 197.1 ks, and the deepest region with 68.8 ks. The key parameters of the \hdha\ project are 
summarized in Tab.~\ref{tab:summary table}.

\begin{figure*}[htbp]
\centering
\includegraphics[width=1.0\textwidth]{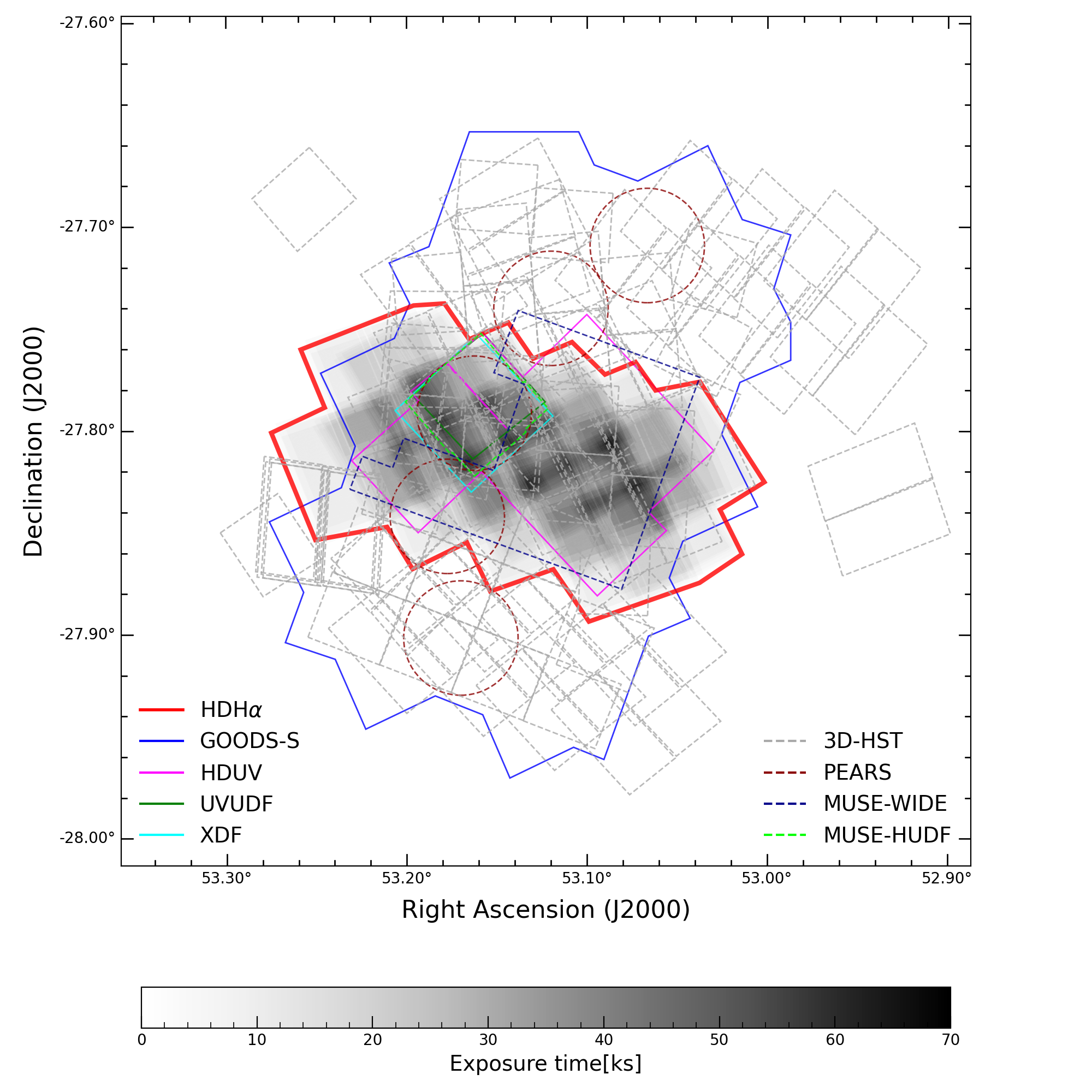}
\caption{The exposure time map of the HDH$\alpha$ and the footprints of other imaging and spectroscopic surveys. The depth of the exposure time map is encoded by the color bar at the bottom. The solid line shows the coverage of broadband imaging surveys such as the GOODS-S \citep[][]{Giavalisco04}, the Hubble Deep UV Legacy Survey \citep[HDUV,][]{Oesch2018}, Hubble Ultraviolet Ultra Deep Field \citep[UVUDF,][]{Teplitz2013}, the Hubble eXtreme Deep Field \citep[XDF,][]{Illingworth2013}, and our narrowband project the \hdha. 
The dashed lines demonstrate the coverages and the pointings of three spectroscopic surveys, i.e., the 3D-HST \citep{Brammer2012, Momcheva2016}, the Probing Evolution And the Reionization Spectroscopically (PEARS) program \citep{Pirzkal2013}, the MUSE-Wide \citep{Urrutia2019} and the MUSE \textit{Hubble} Ultra Deep Field surveys \citep[MUSE-HUDF,][]{Bacon2022}.}
\label{fig:mapHDHa}
\end{figure*}

The narrowband data of the \hdha\ was probed partially by \citet{Finkelstein2011} to study the \lya\ morphology of three LAEs at $z \sim 4.4$. 
However, such narrowband data has never been combined and released as a unified data set. 
In contrast, the broadband images in this field had been reduced and released by
previous programs \citep[e.g., GOODS and HLF;][Illingworth et al. in prep.]{Giavalisco04, Illingworth2016}.
Therefore, a fully reduced narrowband image in this field will increase the legacy value of the GOODS-S field.
The footprints of the \hdha\ project, as well as other large imaging and spectroscopic surveys in the GOODS-S region, are 
presented in Fig. \ref{fig:mapHDHa}.

\begin{deluxetable}{ll}[!h]
\tablecaption{The Summary of the HDH$\alpha$ Project \label{tab:summary table}}
\tablewidth{\linewidth}
\tablecolumns{2}
\tablehead{\colhead{Parameter} & \colhead{Value}}
\startdata
    &  \\
    Position(J2000)& R.A.$=03^{h}32^{m}33^{s}$, Dec$=-27^{\circ}48^{\prime}52\arcsec$ \\
    Area& 86 arcmin$^{2}$ (76.1$^a$ arcmin$^{2}$ )\\
    Camera and Filter& \hst ACS/WFC F658N\\
    Number of Exposures& 214 (212$^b$)\\
    Total Exposure Time& 197,122s (195,690s$^b$)\\
    Number of Orbits & 72 \\
    $5\sigma$ Depth\ & m$_\textrm{F658N}(5\sigma)$ = 25.4\ - 25.8 mag
\enddata
\tablenotetext{a}{The new area after excluding regions with very shallow narrowband exposures. See Section \ref{sec:source detection} for details.}
\tablenotetext{b}{The number of exposures and total exposure time after excluding bad regions with only 2-frame observations. See Section~\ref{subsec:alignment} for details.}
\end{deluxetable}

\subsection{HST's Narrowband Image Processing} \label{sec:data:reduc}

\subsubsection{Data acquisition} 
\label{subsubsec:dq}

We obtain the calibrated \hdha\ images from Mikulski Archive for Space Telescopes (MAST)\footnote{\url{https://archive.stsci.edu/}. All the {\it HST} ACS/WFC F658N data used in this paper can be found in MAST: \dataset[10.17909/v2hh-ns80]{http://dx.doi.org/10.17909/v2hh-ns80}.}. 
These images have been calibrated with the ACS pipeline \textit{calacs}\footnote{See Chapter 3 of the ACS data handbook, \url{https://hst-docs.stsci.edu/acsdhb}}, whose function is briefly summarized here.
The calibration pipeline starts with raw exposures and initially flags bad pixels in the data quality array from a known bad pixel table. Then, bias images and bias levels determined from overscan regions are subtracted. In this step, the bias shift, cross-talk, and striping are also corrected for the full frame ACS/WFC images. After this, the charge transfer efficiency (CTE) degradation is corrected, followed by cosmic ray identification.
Lastly, the dark frame subtraction and the flat field correction are carried out.

We then inspect the quality of the downloaded data visually. Bad quality regions are then masked out. 
In addition, we check two keywords, \texttt{QUALITY} and \texttt{EXPFLAG}, in the image header of the individual frame for the guide star acquisition failure problem. 
All these images pass that check.

\subsubsection{Narrowband Image Registration} \label{subsec:alignment}

Due to the positional uncertainties of guide stars acquired during \hst's observations, the World Coordinate System (WCS) of 
the narrowband images needs to be refined for a precise registration across all images in our data set.
Furthermore, as there are quite a few bright sources but many cosmic rays in an individual narrowband image taken with \hst, 
it is quite difficult to align these images directly. Here we follow the alignment method introduced by \citet{Mack2016}, in which a set of WFC3 narrowband images of Messier 16 were aligned and combined to overcome the alignment problem in the \hdha. 

Our registration procedure begins with drizzling the narrowband images in the same visit and then works on the visit-based narrowband images for further alignment and combination. We use \textit{DrizzlePac}\footnote{\url{https://hst-docs.stsci.edu/drizzpac}}, a suite of tasks including \textit{AstroDrizzle}, \textit{TweakReg} and, \textit{Tweakback} for aligning, distortion-correcting, cosmic-ray cleaning, and combining \hst\ images.
We first combine the individual exposure images within the same visit to obtain a cosmic-ray cleaned, relatively deep image with \textit{AstroDrizzle} \citep[version 3.3.0,][]{Gonzaga2012}. Then we use the \textit{SExtractor} \citep[version 2.19.5,][]{Bertin1996} to identify the sources and create a catalog for each visit-based image. We cross match
our catalog with the reference catalog, which contains bright sources ($\textrm{F606W}<$ 23.5 mag) from the HLF catalog \citep[HLF, ][]{Whitaker2019}. This step is used to remove the residual cosmic rays for each visit. 
Then the WCS solutions for visit-based images are solved by the \textit{TweakReg} (version 1.4.7) task, which is designed to improve the alignments of HST's images.
The WCS fitting results of each visit are shown in Tab.~\ref{tab:alignmentstat}. Finally, these results are
applied to the individual exposure image in each visit by the \textit{Tweakback} (version 0.4.1) task. 

The typical root-mean-square (RMS) value of the alignment residual for visit-based images is $\sim$ 0.3 pixels 
(Tab.~\ref{tab:alignmentstat}). Although it is larger than the value recommended by the ACS data handbook ($\sim$ 0.1 pixels), 
our WCS solution is based on tens of bright sources ($\simeq 10-20$), which is similar to the case of the 3D-HST infrared image \citep[RMS $\sim$ 0.4 pixels for WFC3/IR $F160W$ images,][]{Skelton14}. We reject one visit's data with only two exposures due to a poor WCS solution.

\begin{deluxetable*}{lllllllll}[t]
\tablecaption{The WCS fitting results of individual visits of the HDH$\alpha$.} \label{tab:alignmentstat}
\tablecolumns{11}
\tablehead{\colhead{Visit} & \colhead{Pointing Position} & \colhead{Number of} & \colhead{Total Time of} & \colhead{Shift$_X$} & \colhead{Shift$_Y$}  &  \colhead{RMS$_X$} & \colhead{RMS$_Y$} & \colhead{Number of } \\
\colhead{ID} & \colhead{} & \colhead{ Exposures} & \colhead{Exposures} & \colhead{} & \colhead{} & \colhead{}  & \colhead{} & \colhead{Matched Sources} \\
\colhead{} & \colhead{} & \colhead{} & \colhead{(second)} & \colhead{(pixel)} & \colhead{(pixel)} & \colhead{(pixel)} & \colhead{(pixel)} & \colhead{}}
\startdata
01 & $03^{h}32^{m}46.07^{s} -27^{d}46^{m}33.98^{s}$ & 12 & 8802.0 & 0.96 & -2.32 & 0.36 & 0.39 & 18 \\
02 & $03^{h}32^{m}37.58^{s} -27^{d}47^{m}07.84^{s}$ & 12 & 8802.0 & -0.47 & -0.93 & 0.23 & 0.38 & 11 \\
03 & $03^{h}32^{m}29.09^{s} -27^{d}47^{m}42.19^{s}$ & 12 & 8802.0 & -1.41 & -2.29 & 0.23 & 0.15 & 16 \\
04 & $03^{h}32^{m}20.66^{s} -27^{d}48^{m}17.36^{s}$ & 12 & 8802.0 & -0.12 & 2.43 & 0.40 & 0.35 & 16 \\
05 & $03^{h}32^{m}12.07^{s} -27^{d}48^{m}52.21^{s}$ & 12 & 8802.0 & -0.46 & -0.35 & 0.23 & 0.20 & 20 \\
06 & $03^{h}32^{m}48.92^{s} -27^{d}48^{m}31.08^{s}$ & 12 & 8802.0 & -0.21 & 0.37 & 0.32 & 0.10 & 7 \\
07 & $03^{h}32^{m}40.38^{s} -27^{d}49^{m}05.69^{s}$ & 12 & 8802.0 & -1.59 & -1.25 & 0.17 & 0.44 & 13 \\
09 & $03^{h}32^{m}23.44^{s} -27^{d}50^{m}13.86^{s}$ & 12 & 8802.0 & 0.27 & 2.81 & 0.21 & 0.33 & 24 \\
10 & $03^{h}32^{m}14.82^{s} -27^{d}50^{m}49.56^{s}$ & 12 & 8802.0 & -1.42 & -3.28 & 0.16 & 0.28 & 11 \\
11 & $03^{h}32^{m}45.96^{s} -27^{d}46^{m}29.17^{s}$ & 3 & 2139.0 & 0.54 & -3.02 & 0.23 & 0.19 & 8 \\
12 & $03^{h}32^{m}37.47^{s} -27^{d}47^{m}02.97^{s}$ & 3 & 2139.0 & 1.05 & 0.43 & 0.32 & 0.26 & 7 \\
13 & $03^{h}32^{m}28.99^{s} -27^{d}47^{m}37.37^{s}$ & 3 & 2139.0 & 0.08 & 0.22 & 0.14 & 0.27 & 12 \\
14 & $03^{h}32^{m}20.55^{s} -27^{d}48^{m}12.51^{s}$ & 3 & 2139.0 & 0.17 & 2.21 & 0.37 & 0.16 & 16 \\
15 & $03^{h}32^{m}11.97^{s} -27^{d}48^{m}47.36^{s}$ & 3 & 2139.0 & -0.72 & -0.59 & 0.13 & 0.17 & 16 \\
16 & $03^{h}32^{m}48.82^{s} -27^{d}48^{m}26.31^{s}$ & 3 & 2139.0 & -0.12 & 0.63 & 0.20 & 0.29 & 8 \\
17 & $03^{h}32^{m}40.28^{s} -27^{d}49^{m}00.87^{s}$ & 3 & 2139.0 & 0.35 & 0.57 & 0.29 & 0.52 & 17 \\
18 & $03^{h}32^{m}31.75^{s} -27^{d}49^{m}33.94^{s}$ & 3 & 2139.0 & -2.29 & -6.48 & 0.33 & 0.14 & 6 \\
19 & $03^{h}32^{m}23.33^{s} -27^{d}50^{m}09.05^{s}$ & 3 & 2139.0 & -0.57 & 0.04 & 0.18 & 0.13 & 16 \\
20 & $03^{h}32^{m}14.72^{s} -27^{d}50^{m}44.80^{s}$ & 3 & 2139.0 & -0.80 & -0.12 & 0.39 & 0.18 & 14 \\
25 & $03^{h}32^{m}50.89^{s} -27^{d}46^{m}32.94^{s}$ & 8 & 10964.0 & 1.07 & 1.64 & 0.37 & 0.30 & 9 \\
26 & $03^{h}32^{m}39.24^{s} -27^{d}47^{m}18.52^{s}$ & 8 & 10964.0 & 0.90 & 0.84 & 0.32 & 0.17 & 14 \\
27 & $03^{h}32^{m}27.53^{s} -27^{d}48^{m}04.40^{s}$ & 8 & 10964.0 & 0.37 & -0.85 & 0.26 & 0.22 & 20 \\
28 & $03^{h}32^{m}15.79^{s} -27^{d}48^{m}50.95^{s}$ & 8 & 10964.0 & 0.02 & -1.08 & 0.35 & 0.25 & 23 \\
29 & $03^{h}32^{m}54.82^{s} -27^{d}48^{m}59.81^{s}$ & 8 & 10964.0 & -0.07 & 1.76 & 0.18 & 0.26 & 10 \\
30 & $03^{h}32^{m}43.22^{s} -27^{d}49^{m}43.76^{s}$ & 8 & 10964.0 & -1.13 & -8.73 & 0.27 & 0.26 & 15 \\
31 & $03^{h}32^{m}31.53^{s} -27^{d}50^{m}30.52^{s}$ & 8 & 10964.0 & -0.24 & -0.17 & 0.12 & 0.20 & 23 \\
32 & $03^{h}32^{m}19.75^{s} -27^{d}51^{m}16.53^{s}$ & 8 & 10964.0 & -0.86 & 0.01 & 0.23 & 0.28 & 21 \\
a8 & $03^{h}32^{m}31.48^{s} -27^{d}49^{m}44.10^{s}$ & 3 & 2221.0 & -1.81 & -5.95 & 0.40 & 0.30 & 7 \\
b8 & $03^{h}32^{m}31.99^{s} -27^{d}49^{m}35.61^{s}$ & 4 & 2928.0 & -0.79 & -5.71 & 0.41 & 0.16 & 5 \\
c8 & $03^{h}32^{m}31.96^{s} -27^{d}49^{m}43.71^{s}$ & 3 & 2221.0 & -3.47 & -6.04 & 0.16 & 0.05 & 8 \\
\enddata
\end{deluxetable*}

\subsubsection{Combination} 
\label{sec:com}

The last image processing step combines the well-aligned calibrated images into a final mosaic image. We use 
\textit{AstroDrizzle} to drizzle and fine-tune all narrowband F658N images to the pixel grid of the HLF image. 
The final image has a size of 25000$\times$ 25000 pixels, is located in the GOODS-S field, and has a pixel scale of $0^{\prime \prime}.06$. 
This step's final products of \textit{AstroDrizzle} consist of a science image and a corresponding inverse-variance weight map. 

We also generate the efficient exposure map, which encodes the total exposure time on each pixel, by resampling and combining the efficient exposure maps of different visits with \textit{Swarp} \citep[version 2.38.0,][]{Bertin2010}. 
The efficient exposure map of each visit is one type of weight map generated by \textit{AstroDrizzle}. 

\subsection{Data Quality Check} 
\label{sec:data:quality}

To check the image quality of the \hdha, we conduct tests on the point spread function (PSF), check the astrometry, and probe the distributions of the exposure time and the magnitude depth in this subsection. 

\subsubsection{PSF} 
\label{sec:psf}

The PSF is a basic parameter of an astronomical image. It is also essential for multi-band photometric measurements. 
To check the consistency of fluxes in different bands, we follow previous projects \citep[e.g., 3D-HST, HLF;][]{Skelton14, Whitaker2019} to measure the empirical PSF (ePSF) of the narrowband F658N image of the \hdha\ and four broadband (F435W, F606W, F775W, F814W) images of the HLF, which are used together to search for emission-line galaxies in Section~\ref{sec:HDHaELGs}.

\begin{figure*}[hbtp]
    \centering
    \includegraphics[width= 1.\textwidth]{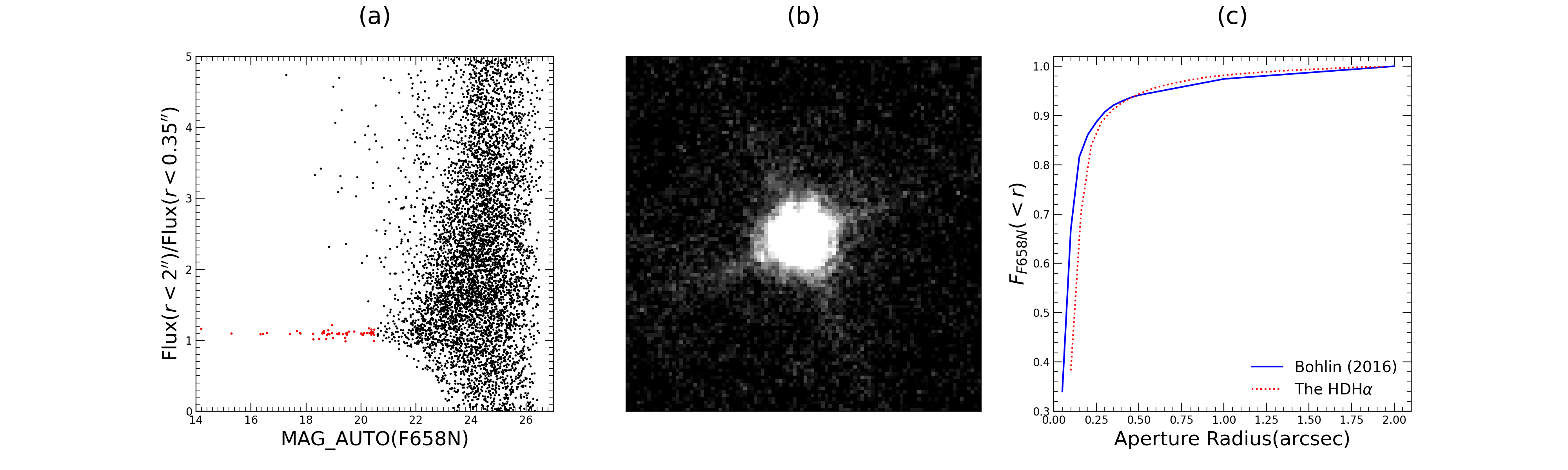}
    \caption{Derivation of the ePSF for the \hdha. 
    (a): the flux ratio in different apertures as a function of the narrowband magnitude. The bright and unsaturated stars are marked in red dots; 
    (b): the ePSF in a $\mathbf{\sim 6^{\prime \prime} \times 6^{\prime \prime}}$ box 
    stacked from bright and unsaturated stars in the \hdha. The diffraction spike structure of the ePSF is clearly shown here; 
    (c): The light growth curves derived from the ePSF (the red dotted line) and from the latest photometric calibration result of HST \citep{Bohlin16} (the blue solid line).}
\label{fig:psf}
\end{figure*}

Following \citet{Skelton14}, we obtain the ePSF of each image by stacking unsaturated bright stars across the mosaic image.  
We describe the method briefly here. 
We select an initial sample of stars from the horizontal sequence on the plot of narrowband flux ratios at different aperture sizes as a function of the Kron aperture \citep{Kron1980} magnitudes, shown in panel (a) of Fig.~\ref{fig:psf}.
This sample is visually inspected to exclude unsuitable objects, such as artifacts, stars affected by nearby neighbors or cosmic rays. 
The number of usable stars varies by band, ranging from a minimum of 31 in F658N to a maximum of 162 in F606W.
For each star, we generate a cutout with the star centered and a size of 100 $\times$ 100 pixels ($\sim 6^{\prime \prime} \times 6^{\prime \prime}$).
Pixels belonging to nearby objects are masked according to the segmentation map,
which encodes whether the pixel belongs to individual objects or the background.
A circular annulus aperture with an inner diameter of $4^{\prime \prime}$ and an outer diameter of $5^{\prime \prime}$ is used to measure the background of each cutout. We then stack all cutouts together to obtain the final ePSF of each image, normalized to the flux enclosed by a circular aperture with a diameter of $4^{\prime \prime}$. We do not consider the positional variation of PSFs of the \hdha\ image. Since the mosaic \hdha\ image comprises \hst's ACS/WFC images at multiple pointings with different orientation angles, the spatial variation of ePSF is averaged to be a small value, as noted by \citet{Skelton14}.

The final ePSF of the \hdha\ image is shown in panel (b) of Fig.~\ref{fig:psf}. Panel (c) of Fig.~\ref{fig:psf} shows 
the comparison of the light growth curves, defined as the flux enclosed by apertures as a function of aperture radius, between the 
latest result of the photometric calibration of ACS/WFC in \citet{Bohlin16} and our derivation. The light growth curves are normalized to 
the aperture flux within a diameter of $4^{\prime \prime}$. The enclosed energy fraction is smaller than expected at a small radius. This problem 
could be caused by \textit{AstroDrizzle} mistakenly flagging the peak of stars as cosmic rays, which is also stated as a common issue of ACS images in previous projects \cite[e.g., the reductions of WFC3/IR F160W images of 3D-HST; See Appendix A  of][]{Skelton14}.

In our evaluation of the enclosed fraction from the final ePSF, we observed a discrepancy at smaller radii, particularly when $r \lesssim 0^{\prime \prime}.3$ (see Panel (c) of Fig.~\ref{fig:psf}). However, at more extensive radii, the results align well with the theoretical expectations. It's worth noting that our primary purpose for the PSF in this research is the aperture correction. Given the close match between the expected PSF and actual measurements at broader apertures, we're confident in the accuracy of our aperture correction across multi-band photometric measurements.

\subsubsection{Astrometry}
\label{sec:astrometric}

\begin{figure*}[hbtp]
    \centering
    \includegraphics[width=\textwidth]{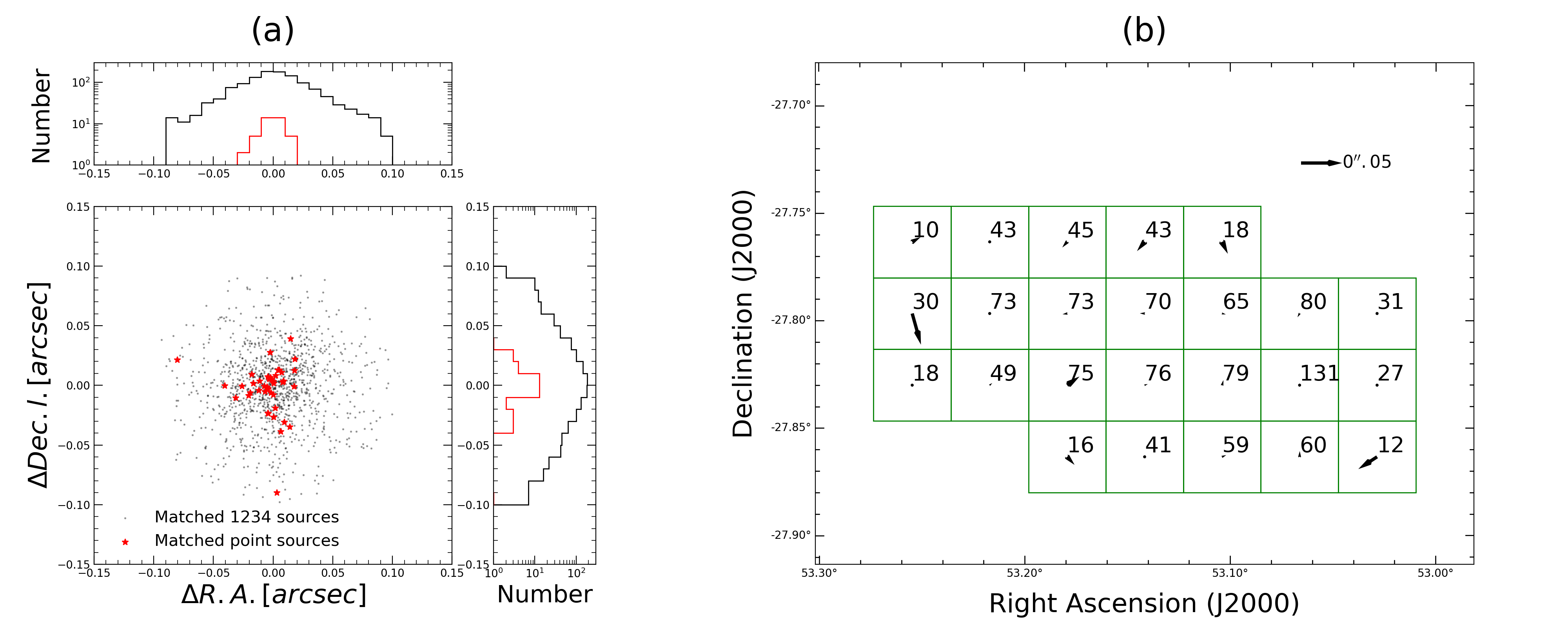}
    \caption{(a): The astrometric residuals calculated from the matched 1234 sources in the \hdha\ project.
    The grey and red histograms in the R.A. and Decl. directions show distributions of the residual offsets for all matched sources and a sub-sample of matched point sources, respectively. 
    (b):The spatial distribution of the astrometric residuals of the matched sources. 
    The panel demonstrates the mean residual shown as a vector in each mesh within a size of $2^{\prime} \times 2^{\prime}$, and the number of matched sources used in each mesh in the calculation.}
    \label{fig:offset_dist}
\end{figure*}

We check the astrometry between our \hdha\ image and the HLF images. By matching the sources with S/N $>10$ in both the \hdha\ catalog and the HLF F606W-band  catalog within a radius of $0^{\prime \prime}.1$, we select 1234 sources for the calculations of the astrometric offsets in the two bands.
We present the scatter map of astrometric residuals and the distribution of offsets in R.A. and Decl. directions. We fit 
the distribution to a Gauss function. The mean offset of the astrometric residuals is at the level of $< 0^{\prime \prime}.002$, 
and the 1 $\sigma$ values on R.A. and Decl. are $\mathrm{0^{\prime \prime}.033}$ and $\mathrm{0^{\prime \prime}.032}$, respectively.
The offset corresponds to a spatial difference of less than 0.1 pixel (See panel (a) of Fig.~\ref{fig:offset_dist}), which is quite good when considering the positional uncertainties of sources in the \hdha. 
Panel (b) of Fig.~\ref{fig:offset_dist} shows the astrometric residuals at different positions using the same data as panel (a). 
To explore the spatial variation of the alignment, we check the residuals at different areas of the \hdha, which are composed of different meshes of the \hdha\ with each mesh's size of 2000 pixels $\times$ 2000 pixels ($2^{\prime} \times 2^{\prime}$).
We present the average value and the direction (vector) of sources in each mesh in panel (b) of Fig.~\ref{fig:offset_dist}.
The mean residuals are small across the image, meaning the residual offsets of sources in 
each mesh are distributed randomly. Therefore, there is no large systematic offset exists. 

\subsubsection{Exposure Time and Depth}
\label{sec:depth}

We measure the differential and cumulative distributions of the exposure time and the 5$\sigma$ limiting magnitude of the \hdha\ image. 
The distributions of exposure time are measured based on the exposure time map generated in Section~\ref{sec:com}. 

\begin{figure}[htbp]
    \centering
    \includegraphics[width=0.5\textwidth]{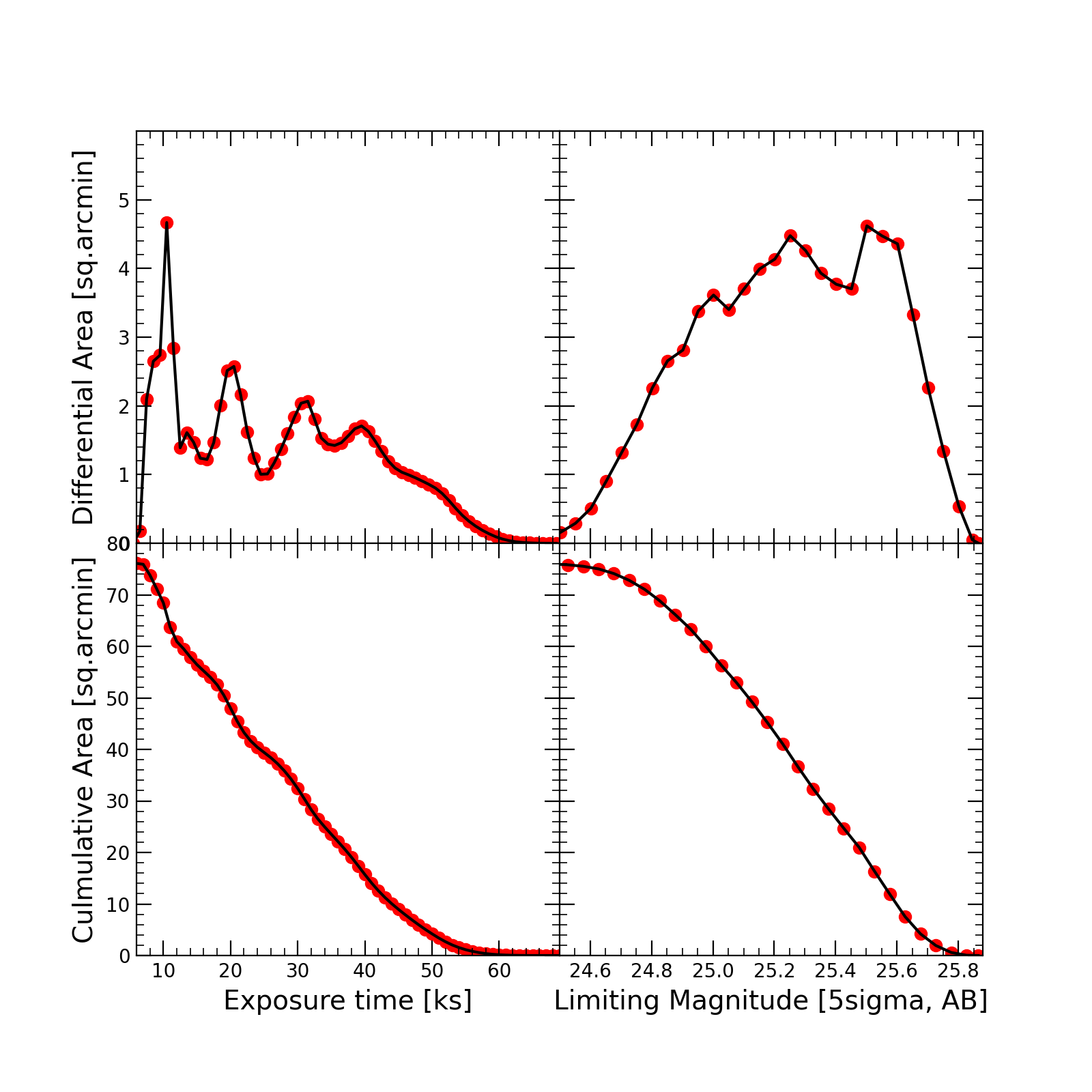}
    \caption{The differential (top row) and cumulative (bottom row) distributions of the effective exposure time (left column) and the 5$\sigma$ depth magnitude (right column) in the \hdha\ project.}
\label{fig:exposure_dict}
\end{figure}

We use the same way as the previous work \citep[e.g. HDUV,][]{Oesch2018} to estimate the depth of the \hdha.
We measure the distribution of 5$\sigma$ limiting magnitude by dividing the images into meshes with a size of $\rm 128 \ pixels \times 128 \ pixels$ ($7^{\prime \prime}.68\times7^{\prime \prime}.68$).
For each mesh, the 5$\sigma$ limiting magnitude is measured by randomly placing empty apertures with a diameter of 0$^{\prime \prime}$.7 in the background region and estimating the standard deviation of the fluxes in these empty apertures with the zero-point magnitude of 22.75.
The zero-point magnitude here is determined with two keywords of \texttt{PHOTFLAM} and \texttt{PHOTPLAM}, which are generated by \textit{AstroDrizzle} in the image header of the mosaic \hdha\ image\footnote{$\rm ZP_{AB} = -2.5 \times log10(\texttt{PHOTFLAM})-5 \times log10(\texttt{PHOTPLAM})-2.408$. \\
\href{https://www.stsci.edu/hst/instrumentation/acs/data-analysis/zeropoints}{https://www.stsci.edu/hst/instrumentation/acs/data-analysis/zeropoints} 
}. 
The results are shown in Fig. \ref{fig:exposure_dict}.
In the deepest region, the 5$\sigma$ limiting magnitude of \hdha\ is 25.8.

\begin{figure}
    \centering
    \includegraphics[width = 0.5 \textwidth]{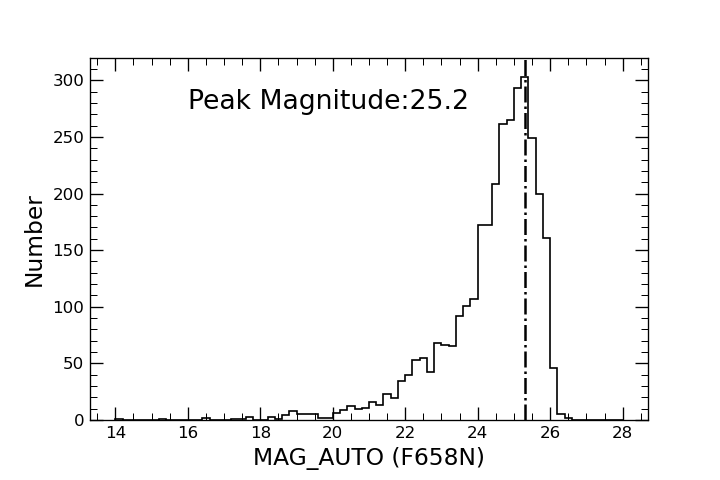}
    \caption{The magnitude distribution of objects with S/N $\geq$ 5 in the mosaic F658N image of the \hdha\ project.
    The peak magnitude is around 25.2 mag here, compared to 24.8 mag in \citet{Finkelstein2011}.}
    \label{fig:hist}
\end{figure}

We also check the depth of the \hdha\ by using the same method as \citet{Finkelstein2011}, which is summarized as below. 
\citet{Finkelstein2011} checked the distribution of all detected sources, as well as the statistics of non-detected sources (in empty apertures) in section 23 of GOODS-S \citep[version 2.0\footnote{\href{https://archive.stsci.edu/prepds/goods/}{https://archive.stsci.edu/prepds/goods/}},][]{Giavalisco04}. Three types of limiting magnitudes are defined here, the peak magnitude, the detection limit magnitude, and the 5$\sigma$ depth magnitude.
The peak of the magnitude distribution of detected sources defines the peak magnitude. At the 50\% peak value in the fainter-end of the distribution, the corresponding magnitude is defined as the detection limit magnitude. The 5$\sigma$ depth magnitude is derived by using the empty aperture method. 
The distributions of the detected sources in the \hdha\ are presented in Fig.~\ref{fig:hist}. The statistics of non-detected sources in the \hdha\ are applied in the same region as that of \citet{Finkelstein2011}.
We present a comparison of the three kinds of limiting magnitudes from ours and those in \citet{Finkelstein2011} in Tab. \ref{tab:depths}.
All three limiting magnitudes indicate that the \hdha\ image is about $\sim$0.4$-$0.5 mag deeper than that of \citet{Finkelstein2011}.

\begin{deluxetable}{lll} \label{tab:depths}
\tablecaption{Depths of the $F658N$ images in the GOODS-S field derived by \cite{Finkelstein2011} and this work.}
\tablewidth{\linewidth}
\tablecolumns{3}
\tablehead{\colhead{} & \colhead{Finkelstein+2011} & \colhead{This work}}
\startdata
Peak magnitude & 24.8 mag & 25.2 mag\\ 
Detection limit magnitude & 25.3 mag & 25.8 mag\\
5$\sigma$ depth magnitude & 25.0 mag & 25.4 mag\\
Number of detected sources & 3081 & 3960\\
Median exposure Time  & $\sim$11ks & $\sim$22ks\\
\enddata
\end{deluxetable}

\section{Photometric and spectroscopic data in \hdha}
\label{sec:photo}

\subsection{Narrowband Source Detection}
\label{sec:source detection}

Since a few high redshift objects would show extended structures in their emission lines, we use the source detection strategy similar to that of the CANDELS program \citep[][]{Guo2013,Galametz2013}, which applied the cold mode and the hot mode with \textit{SExtractor}. 
The cold mode is designed to detect the bright/extended objects in images without over-deblending, while the hot mode is designed to detect the faint/point objects,  
especially for objects at high redshift. The parameters of the cold/hot modes used for the \hdha\ are listed in Tab.~\ref{tab:param_phot}. 

The \hdha\ narrowband image and the RMS map are used as the detection image and the weight map in \textit{SExtractor}, respectively. The cold and hot modes are applied to generate corresponding catalogs. There are 7466 and 7520 sources detected in the cold and hot modes, respectively.
All objects detected in the cold mode are kept. Objects were detected in the hot mode but not in the Kron aperture.
There are 7762 objects in the combined catalog.

Note that the cosmic rays cleaning process of \textit{AstroDrizzle} in the edge region of the \hdha\ image might fail because of too few frames covered in these areas. This could bring in the remaining cosmic rays in the combined
catalog. We require a minimal effective exposure time cut of 10\% of the deepest exposure time to exclude these artifacts. The effective exposure time of each source is defined as the average exposure time of pixels that are enclosed in the photometric aperture. This step successfully filters out fake sources in the edge region. There are 3960 objects in an effective area of 76.1 arcmin$^{2}$.

\begin{deluxetable}{ccc}
\tablecaption{Parameters for the source detection process in the \hdha\ project.}\label{tab:param_phot}
\tablewidth{\linewidth}
\tablecolumns{3}
\tablehead{\colhead{} & \colhead{Cold Mode} & \colhead{Hot Mode}}
\startdata
DETECT\_THRESH & 1.2 & 1.0 \\
ANALYSIS\_THRESH & 5.0 & 1.0 \\ 
FILTER & tophat\_5.0\_5x5 & gauss\_4.0\_7x7\\
DETECT\_MINAREA & 5.0 & 8.0 \\
DEBLEND\_NTHRESH & 32 & 64 \\
DEBLEND\_MINCONT & 0.1 & 0.001 \\
BACK\_SIZE & 128 & 64 \\
BACK\_FILTERSIZE & 9 & 5\\
BACKPHOTO\_THICK & 100 & 48 \\
\enddata
\end{deluxetable}

\subsection{Multi-band Photometric Data} \label{subsec:photometry}
In order to select emission-line galaxies in the HDH$\alpha$, we use four additional broadband images in the HLF (ACS/WFC F435W, F606W, 
F775W and F814W). The transmission curves of these filters are shown in Fig.~\ref{fig:throughput}.
   
We run \textit{SExtractor} in the dual-image mode to measure the aperture fluxes for the HDH$\alpha$ narrowband image and the HLF broadband images, respectively, with the cold-hot strategy described in Section~\ref{sec:source detection}. We use the \hdha\ image as the detection image.
The aperture photometry is calculated within a circle with a diameter of $0^{\prime \prime}.7$.
The aperture correction in each band is applied based on the light growth curve of ePSF we generated in the corresponding band. 
Then the total flux is calculated as:
\begin{equation}
    f_{X,tot} = f_{X, aper} \frac{f_{X}(r_{tot})}{f_{X}(r < r_{aper})}
\end{equation}
where X denotes the filter (i.e., F435W, F606W, F775W, F814W or F658N), the $f_{X, aper}$ is the flux encircled in the aperture, 
and the ratio of $f_{X}(r_{tot})$/$f_{X}(r< r_{aper})$ is measured from the ePSF we generated.
 
The zero-point magnitude of the \hdha\ image is described in Section~\ref{sec:depth}. As for the HLF broadband data, we use the zero-point magnitude provided by the HLF team. We convert the flux after the aperture correction into magnitude using this zero-point. We do not apply any further zero-point correction, which is expected to be very small ($< 0.1 \rm mag$) according to the previous work
\citep[e.g., see Appendix section of][]{Guo2013,Skelton14}.

\subsection{Broadband-over-narrowband Color Offset}
\label{sec:colorzero}

Because the center wavelength of the narrowband filter does not match that of the broadband filter, the corresponding broadband-over-narrowband color of continuum-dominated sources with a non-flat slope would not have a value around zero. We first demonstrate this phenomenon through the calculation.
We define the observer-frame equivalent width of an emission line as 
\begin{equation}\label{eq:ew}
   \mathrm{EW_{obs}} = \Delta \lambda_{NB} \frac{f_{\lambda, NB} - f_{\lambda, BB}}{f_{\lambda, BB}-f_{\lambda, NB}(\Delta \lambda_{NB}/\Delta \lambda_{BB})},
\end{equation}
where the $f_{\lambda, NB}$ and $f_{\lambda, BB}$ are the flux densities in the unit of $\mathrm{erg\ cm^2\ s^{-1}\ \AA^{-1}}$ of the broadband and the narrowband, respectively. The $\Delta \lambda_{NB}$ and $\Delta \lambda_{BB}$ are the widths (FWHMs) of the corresponding filter, and the center wavelengths of the narrowband filter and the broadband filter are 6584 \AA\ and 5907 \AA, respectively.
With Eqn. \ref{eq:ew},
we derive the broadband-over-narrowband color of a continuum-dominated source (EW$_\textrm{obs}$ = 0) as 0.24 mag.

We then use objects with spectroscopic redshifts whose continuum is covered by the F658N narrowband filter to check the broadband-over-narrowband color offset. 
The redshift information is from Section~\ref{sec:zelgs} and the emission lines are listed in Tab.~\ref{tab:elgzrange}. 
Only the bright sources with S/N $\geq$ 10 in narrowband and broadband images are used in this procedure.
We adopt the median after the sigma-clipping to remove the impact of the remaining outliers. 
The sigma-clipped median value and one sigma value of the colors are $\sim 0.27$ mag and $\sigma \sim 0.14$, respectively, which agree with the calculated result above.

\begin{figure}[th]
    \centering
    \includegraphics[width=0.48\textwidth]{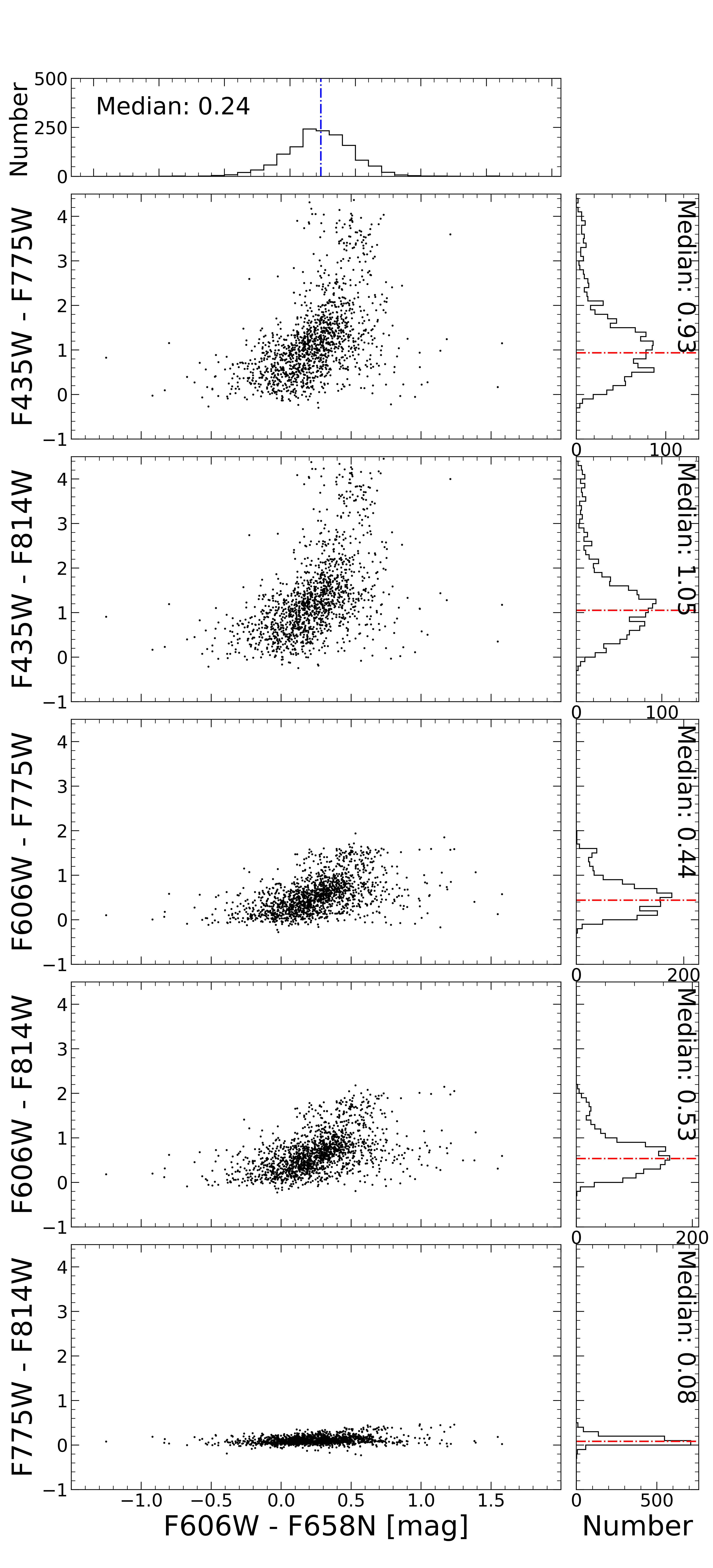}    
    \caption{The relations between the broadband-over-narrowband color and different broadband colors. We investigate the relations using sources with S/N $>$ 10 in the corresponding broadbands as well as only the continuum covered by the F658N narrowband.
    Most of the sources in our field have relatively red colors, which are reflected by the positive broadband colors and the increasing differences in the central wavelengths of the two corresponding filters.}
    \label{fig:color_distri}
\end{figure}

To further probe how the broadband color (or the continuum slope) affects the broadband-over-narrowband color offset, we follow \citet{Sobral13} to investigate the relations between the broadband-over-narrowband color and the broadband color for sources with a bright continuum. 
The relations between the broadband-over-narrowband color and five different broadband colors are presented in Fig.~\ref{fig:color_distri}.
The median values of the distributions are calculated with the sigma-clipping method. In Fig.~\ref{fig:color_distri}, we can see that the effect of broadband colors can be ignored when estimating the broadband-over-narrowband color offset.

\begin{deluxetable}{lll}
    \tablecaption{Redshift ranges of line emitters whose emission line could be covered by the F658N narrowband filter.}
    \tablewidth{\linewidth}
    \tablecolumns{3}
    \tablehead{\colhead{Emission line} & \colhead{Rest-frame wavelength} & \colhead{Redshift range}}
    \startdata
        \lya & 1215.67\AA & [4.370, 4.463]\\
        O\,{\sc i} & 1302.168\AA & [4.013, 4.100]\\
        N\,{\sc iv}] & 1486.496\AA & [3.392, 3.468]\\
        C\,{\sc iv} & 1548.187\AA/1550.772\AA & [3.210, 3.290]\\
        C\,{\sc III}] & 1908.734 \AA & [2.420, 2.479]\\
        Mg\,{\sc ii}] & 2795.528\AA/2802.705\AA & [1.329, 1.376]\\
        \oii & 3726.032\AA /3728.815\AA & [0.751, 0.782] \\
        \heii & 4685.71\AA & [0.393, 0.417]\\
        \hb & 4861.333\AA & [0.343, 0.366]\\
        \oiii & 4958.911\AA / 5006.843\AA & [0.304, 0.339] \\
        \ha & 6562.819\AA & [0, 0.012]\\
    \enddata
    \label{tab:elgzrange}
    \tablecomments{The filter width used to calculate the redshift range is the full width at the 5\% maximum (wavelength range: 6528\AA-6641\AA). The wavelength values of the emission lines are from \citet{NIST_ASD}.}
\end{deluxetable}

\subsection{Spectroscopic data}
\label{subsec:speccat}
The enormous spectroscopic resources in the GOODS-S field help us to validate the redshifts of ELG candidates selected in the \hdha. To obtain the spectroscopic redshifts for sources identified with \hdha, the following public spectroscopic catalogs are used, including (1)The compilation of the spectroscopic catalog of VLT, which assembles the major results of large spectroscopic surveys between 2001-2012 \citep[][]{Vanzella2008, Le_F2005, Szokoly2004, Croom2001, Dickinson2004, van2004, Bunker2003, stanway2004, Doherty2005, Cristiani2000, Strolger2004, stanway2004, Popesso2009, Balestra2010, Mignoli2005, Ravikumar2007, Silverman2010, Kurk2013}; (2)The VANDELS DR4 catalog based on VANDELS spectroscopic survey \citep{Garilli2021}; (3)The MUSE-Wide catalog \citep{Urrutia2019}; (4) The catalog of the MUSE-HUDF survey \citep{Bacon2022}.

\begin{deluxetable}{llll}[t] \label{tab:spec-table}
    \tablecaption{Summary of spectroscopic catalogs in the GOODS-S region used in this work.}
    \tablewidth{\linewidth}
    \tablecolumns{3}
    \tablehead{\colhead{  } & \colhead{Total} & \colhead{Number} & Number \\ Catalog name &  & of sources & of matched \\  & Number & in \hdha &  with \hdha }
    \startdata
        The compilation catalog$^{1}$ & 7146 & 2115 & 1095\\
        VANDELS DR4 catalog$^{2}$ & 1017 & 252 & 97\\
        MUSE-Wide catalog$^{3}$ & 1602 & 1422 & 709\\
        MUSE-HUDF catalog$^{4}$ & 2221 & 2221 & 426\\
    \enddata
    \tablenotetext{1}{\href{https://www.eso.org/sci/activities/garching/projects/goods/MasterSpectroscopy.html}{https://www.eso.org/sci/activities/garching/projects/goods/\\MasterSpectroscopy.html}}
    \tablenotetext{2}{\href{http://vandels.inaf.it/dr4.html}{http://vandels.inaf.it/dr4.html}}
     \tablenotetext{3}{\href{https://musewide.aip.de/dr1/photometric_catalog}{https://musewide.aip.de/dr1/photometric\_catalog}}
    \tablenotetext{4}{\href{https://amused.univ-lyon1.fr/project/UDF/HUDF/}{https://amused.univ-lyon1.fr/project/UDF/HUDF/}}
\end{deluxetable}

We also note that there are grism catalogs of HST based on the PEARS program \citep[PI: Sangeeta Malhotra, Program ID: 10530,][]{Pirzkal2013} and the 3D-HST program\citep[PI: Pieter van Dokkum, Program ID: 12177, 12328,][]{Brammer2012, Momcheva2016}.
There are no \hdha\ ELG or LAE candidates matched with sources in the PEARS catalog. For the 3D-HST catalog, we found that 37 sources in \hdha\ confirmed at $z = 0$ from ground-based spectroscopic surveys show $z>0.1$ in the 3D-HST grism catalog.
Therefore, we do not use these two catalogs in this work.

We plot the coverage of the MUSE-Wide, the MUSE-HUDF, the PEARS, and the 3D-HST, along with several imaging data sets in the GOODS-S filed in Fig~\ref{fig:mapHDHa}. The total number of sources in various catalogs and in the \hdha\ coverage are listed in Tab.~\ref{tab:spec-table}. We perform a cross matching between the redshift catalogs described above and the \hdha\ catalog with a matching radius of $\boldmath 0^{\prime \prime}.6 \unboldmath$. There are 1,799 sources that have at least one detection in these spectroscopic catalogs. For those sources with more than one redshifts in the spectroscopic catalogs, their redshifts are chosen based on the spectral resolution and quality of the corresponding catalogs. 

\section{Emission-line Galaxies in \hdha}
\label{sec:HDHaELGs}

In this section, we describe our selection procedure for ELGs with the \hdha\ data. 
The selection criteria of ELGs are discussed in Section~\ref{subsec:sele}.
We discuss the confirmed ELGs in Section~\ref{sec:zelgs}.
Finally, the depths of emission lines are discussed in Section~\ref{sec:linedepth}.

\subsection{ELGs' selection criteria}
\label{subsec:sele}

\hdha\ ELGs should have a strong emission line covered by the F658N narrowband filter, thus can be selected by their broadband-over-narrowband color excess (F606W-F658N) to ensure the existence of a strong emission line, such as \ha\ at $z\sim$0.0, \oiii\ at $z\sim$ 0.3, \oii\ at $z\sim$ 0.8, or \lya\ at $z\sim$ 4.4 (see Tab.~\ref{tab:elgzrange} for details).
Besides the color excess, we also require a line significance parameter on the color excess defined as $\Sigma = \frac{BB - NB}{\sqrt{\Delta BB^2 + \Delta NB^2}}$, where BB and NB represent the magnitudes in broadband and narrowband, and $\Delta BB$ and $\Delta NB$ are the errors of magnitude in broadband and narrowband, respectively. 

Since all \hdha\ sources are covered by the F435W band, according to the Lyman-break technique the detections and non-detections of F435W would divide the redshifts of \hdha\ sources at $z<4$ and $z\gtrsim4$, respectively. 
For the latter case, ELGs at $z\gtrsim4$ selected with \hdha\ would be LAEs at $z\sim$ 4.4. 
We discuss the two samples separately below.

\begin{figure*}
    \centering
    \includegraphics*[width=1\textwidth]{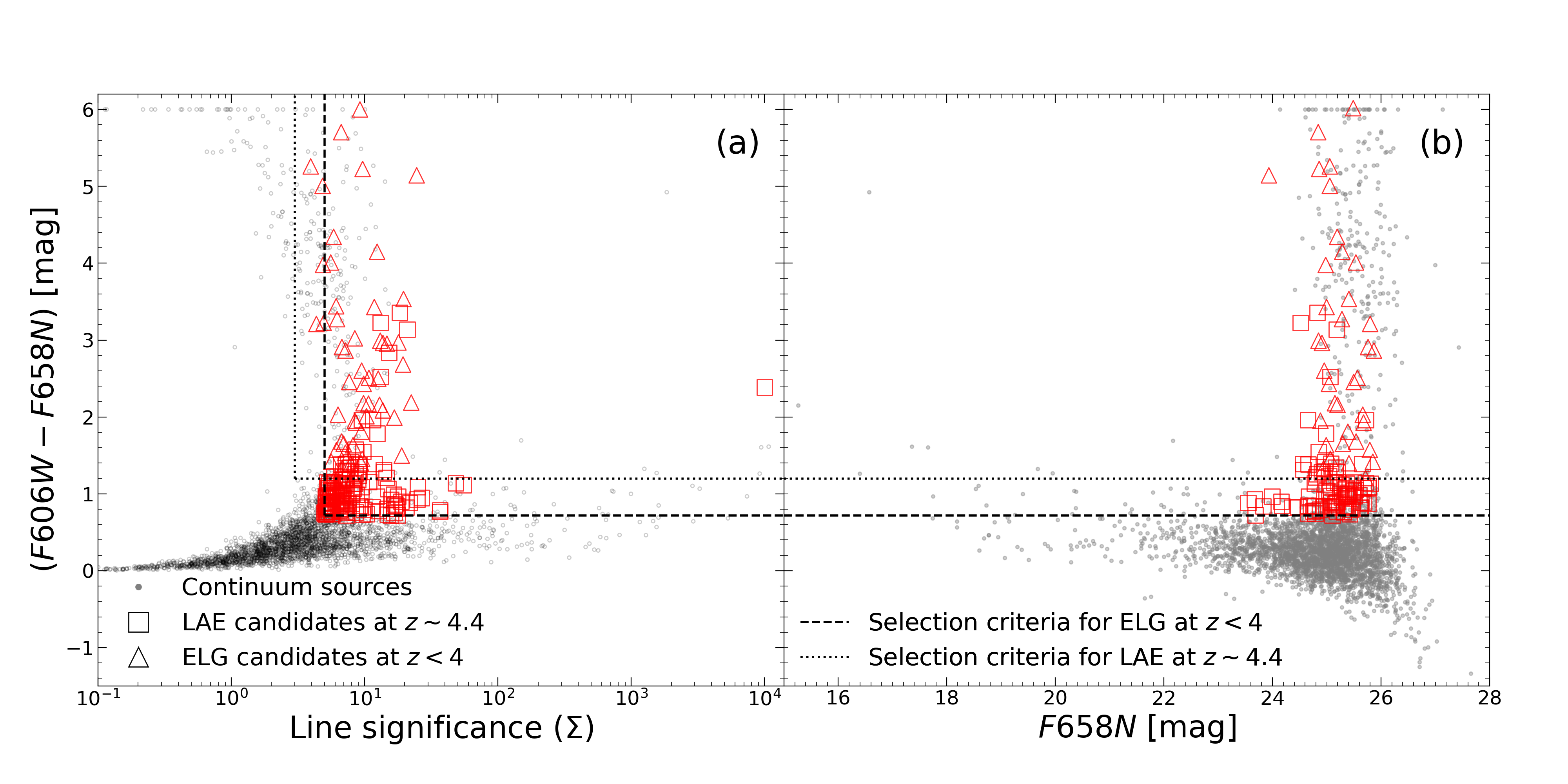}
    \caption{(a): The color and the line significance ($\Sigma$) diagram for the continuum sources (grey dots), 156 ELG candidates (squares), and 53 LAE candidates (triangles) in \hdha. The continuum sources refer to those not selected as neither ELG candidates at $z<4$ nor LAE candidates at $z\sim 4.4$. The dashed line and dotted line represent the color and line significance cuts used for the selections of ELG candidates at $z<4$ and LAE candidates at $z\sim$ 4.4, respectively (see Section~\ref{subsubsec:elg} and Section~\ref{subsubsec:lae}).
    (b): The color-magnitude diagram of the continuum sources, ELG candidates at $z<4$, and LAE candidates at $z \sim 4.4$ in \hdha.
    The symbols are the same as in panel (a).
    Colors of F606W-F658N $>6$ are set to $= 6$ for display only.
    The two horizontal lines mark the color cuts used in the selection criteria for ELGs at $z<4$ and  LAEs at $z\sim4.4$, which correspond to a rest frame EW ${\sim 30}$ \AA\ at $z\sim0.77$ and ${\sim 25}$  \AA\ at $z\sim4.4$, respectively.}
    \label{fig:color_sig_cmd}
\end{figure*}

\subsubsection{ELG candidates at $z<4$}
\label{subsubsec:elg}

We select the ELG candidates at $z<4$ using the selection criteria below, 
which include an S/N cut on the narrow-band detection, a color cut corresponding to the equivalent-width cut of an emission line, and a significance of the color excess. These criteria are similar to those of the ground-based ELG surveys \citep[e.g.,][]{Sobral2015, Khostovan2015, Khostovan2020}.
To further exclude cosmic rays or artifacts, as well as ensure the redshifts at $z<4$, we require sources with reliable signals in the bluer broadband image (here the F435W image). 
The following selection criteria are applied to select ELG candidates at $z<4$:
\begin{eqnarray}
S/N_{\rm F658N} & > & 5.0, \\\nonumber
F606W -  F658N & > & 0.72, \\\nonumber
\Sigma = \frac{F606W -  F658N}{\sqrt{\Delta {F606W}^2 + \Delta {F658N}^2}} & > & 5.0, \\\nonumber
S/N_{\rm F435W} & > & 5.0,  \label{eq:elg}
\end{eqnarray}
where the $S/N_{\rm F658N}$ and the $S/N_{\rm F435W}$ are the signal-to-noise ratio of objects in the F658N image and the F435W image, respectively.
The color cut is set to 0.72, which corresponds to observer-frame equivalent widths of 52.5 \AA, which is the same as the previous survey \citep{Khostovan2020}. This color cut corresponds to rest-frame equivalent widths of 52.5 \AA\
at $z\sim 0.0$, 39.8 \AA\ at $z \sim 0.32$, and 30.0 \AA\ at $z \sim 0.77$. The significance of the color excess $\Sigma$ is often set to 3 according to previous studies \citep[e.g.,][]{Sobral2015, Khostovan2015, Khostovan2020}. Here we set this to a value of 5 to exclude more contamination indicated by the spectroscopic data (see Sec. \ref{sec:zelg:comelg} for details).

There are 212 ELG candidates at $z<4$ selected with the above criteria. After the visual inspection, 4 objects are excluded because of the cosmic rays or artifacts.
The remaining 208 ELG candidates should include \ha\ emitters at $z\sim$0.0, \oiii\ emitters at $z\sim$ 0.3, \oii\ emitters at $z\sim$ 0.8, and other contamination. Note that Galactic stars could be included in the sample of \ha\ emitters. 

To exclude stars in the sample of ELG candidates, especially in the \ha\ emitters, we utilize the star selection method of \citet{Dahlen2010}, which is based on the F435W-J color and J-IRAC1 color. We select 52 objects using this method. 27 of the 52 objects have available redshifts, and all these 27 objects are identified as stars according to their spectra. Furthermore, these 27 stars are all the \hdha\ ELG candidates with available redshifts at $z\sim$ 0. 
Therefore, we exclude all the 52 objects from the ELG candidates at $z<4$ (see the Appendix section for details). 

Finally, we select 156 objects as ELG candidates at $z<4$ in \hdha. The sample should include strong ELGs, such as [O{\sc iii}] emitters at $z\sim $ 0.32 and [O{\sc ii}] emitters at $z\sim $ 0.77.
All these ELG candidates at $z<4$ are marked with squares in the color-line significance($\Sigma$) diagram and the color-magnitude diagram (see panels (a) and (b) of Fig.~\ref{fig:color_sig_cmd}).

\subsubsection{LAE candidates at $z\sim 4.4$}
\label{subsubsec:lae}

The selection criteria for LAEs at $z\sim$ 4.4 are summarized as below, including an S/N cut on the narrow-band detection, a color cut corresponding to the equivalent-width cut of the \lya\ line, a line significance parameter of the color excess, and a blue band  (F435W here) non-detection. These criteria are similar to those of the ground-based LAE surveys \citep[e.g.,][]{Rhoads2000, Finkelstein2008}.
We require a red band (F775W or F814W) detection to further exclude cosmic ray events and other artifacts.

The following selection criteria are applied to select LAE candidates:
\begin{eqnarray}
S/N_{\rm F658N\ Auto} > 5.0 \quad \|\quad S/N_{\rm F658N\ Aper} & > & 5.0, \\\nonumber
 F606W - F658N & > & 1.2, \\\nonumber
\Sigma = \frac{F606W -  F658N}{\sqrt{\Delta {F606W}^2 + \Delta {F658N}^2}} & > & 3.0, \\\nonumber
S/N_{\rm F435W} & < & 3.0,  \\\nonumber
S/N_{\rm F775W}  >  3.5  \quad \|\quad  S/N_{\rm F814W} & > & 3.5, 
\end{eqnarray}
where the $S/N_{\rm F658N\ Aper}$ and $S/N_{\rm F658N\ Auto}$ are the signal-to-noise ratios in the F658N band within an aperture of diameter $0^{\prime \prime}.7$ and a Kron aperture, respectively. The $S/N_{\rm F775W}$ and $S/N_{\rm F814W}$ are the signal-to-noise ratios of objects in F775W and F814W bands within an aperture of diameter $0^{\prime \prime}.7$, respectively. The $S/N_{\rm F435W}$ is measured using an aperture of diameter $0^{\prime \prime}.5$ to avoid the contamination from nearby objects. The color threshold is set to 1.2, corresponding to rest-frame equivalent width of $\sim 25$ \AA\  at $z = 4.4$. We require that the LAEs cannot be detected in the F435W band because LAEs at $z=4.4$ should be drop-outs at the F435W band due to their Lyman break. We also require the sources to be detected in the F775W or F814W band to exclude false detections. 

We selected 66 LAE candidates with the above criteria. After the visual inspection, there are 53 LAE candidates left, which are marked as triangles in the color-line significance($\Sigma$) diagram and the color-magnitude diagram (see panels (a) and (b) of Fig.~\ref{fig:color_sig_cmd}).

\subsection{Spectroscopic confirmation of \hdha\ ELGs}
\label{sec:zelgs}

\begin{figure*}[ht]
    \centering
    \includegraphics[width=\textwidth]{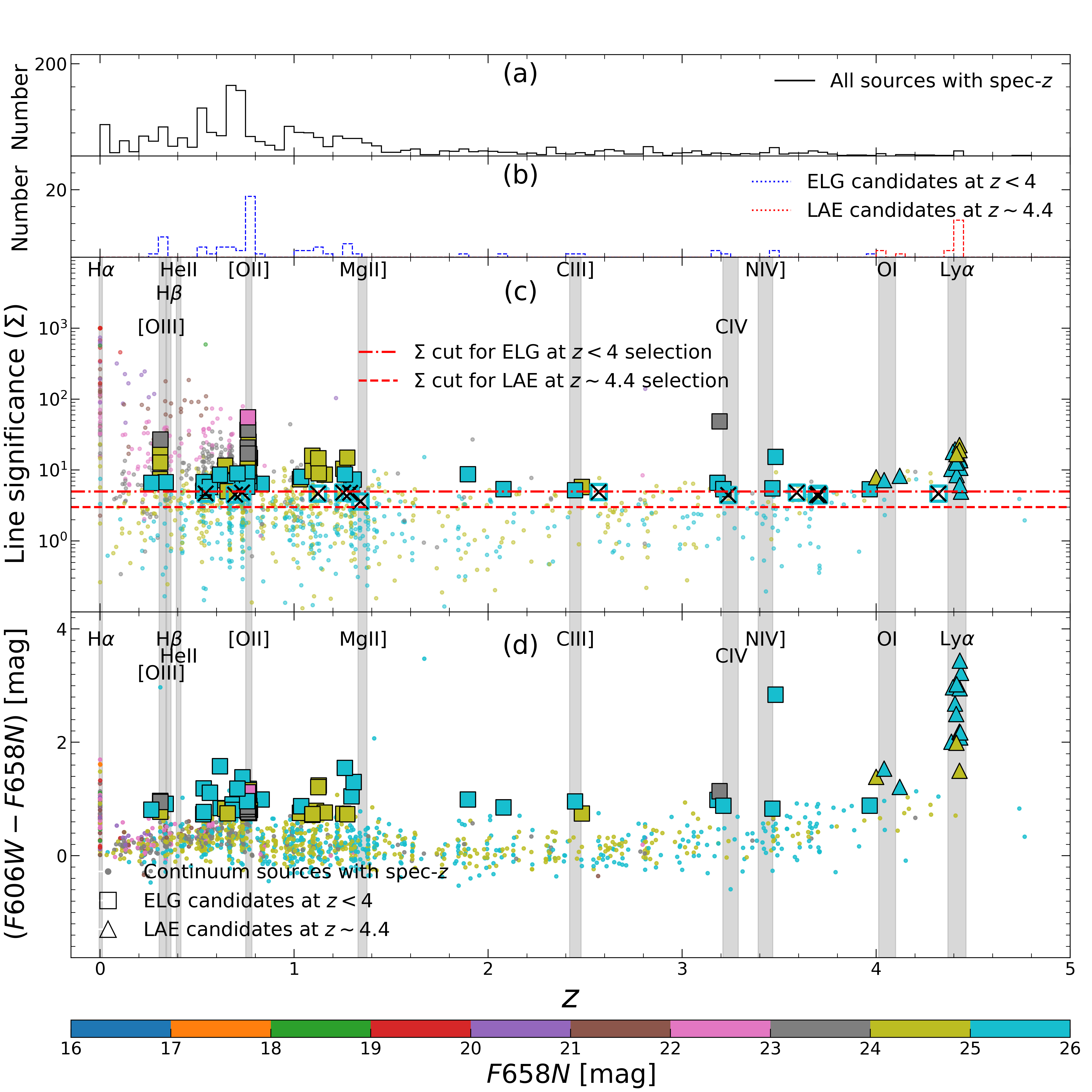}
    \caption{
    (a): The redshift distribution of all \hdha\ sources with available redshifts.
    (b): The redshift distributions of ELG candidates at $z<4$ and LAE candidates $z \sim 4.4$ with available redshifts in the \hdha, which are marked by the blue dashed and the red dashed histograms, respectively.
    (c): The line significance ($\Sigma$) of the color excess and redshift diagram of continuum sources, ELG candidates at $z<4$, and LAE candidates $z\sim 4.4$ in the \hdha.
    The grey vertical regions mark the redshift ranges of the corresponding emission lines. 
    The symbols of ELG candidates at $z<4$ and LAE candidates at $z\sim 4.4$ are similar to those in Fig. \ref{fig:color_sig_cmd}.
    The open squares with cross marks represent sources with F658N-F606W color $>$ 0.72 but line significance $\Sigma<5$, which are not valid ELGs based on their redshifts.
    The symbols' color is encoded with their narrowband magnitudes shown in the bottom color bar. 
    We can see that fainter sources tend to have smaller significance values of their color excesses.
    (d): The color-redshift diagram of continuum sources, ELG candidates at $z<4$, and LAE candidates at $z \sim 4.4$ with available redshifts.
    The symbols are similar to those in panel (c).
    In this panel, sources fainter than 26 mag in F658N are excluded.
    }
    \label{fig:the_redshift_color}
\end{figure*}

In this section, we discuss the selected ELG candidates with available spectroscopic redshifts, including \oiii\ emitters at $z\sim$ 0.3, \oii\ emitters at $z\sim$ 0.8, and LAEs at $z \sim 4.4$. 
In Fig.~\ref{fig:the_redshift_color}, we present the redshift distribution and the color-redshift diagram for all sources with spectroscopic redshifts, ELG candidates at $z<4$, and LAE candidates.
As we introduced in Section~\ref{sec:intro}, the narrowband excess method is only sensitive to galaxies at specific redshifts, where their emission lines are redshifted to the bandpass of the F658N filter. Here we give the expected redshift ranges of 9 emission line emitters in Tab.~\ref{tab:elgzrange}. These emitters' redshift ranges are marked as grey regions in Fig.~\ref{fig:the_redshift_color}.

\subsubsection{Confirmed ELGs at $z<4$}
\label{sec:zelg:comelg}

Among the 156 ELG candidates at $z<4$ selected with \hdha, 61 are matched with the spectroscopic catalogs introduced in Section~\ref{subsec:speccat}. 
We classify these ELG candidates further by matching the redshifts ranges of the corresponding lines in Tab. \ref{tab:elgzrange} and by inspecting their spectra.
In the following paragraphs, we separate the sample into sub-samples at $z<1$ and at $1\lesssim z <4$ for clear discussions.

We have cross-matched 38 ELG candidates with spectroscopic redshifts at $z<1$. Based on their redshifts, we classify 6 objects as \oiii\ emitters and 18 objects as \oii\ emitters. Through visual inspection of their spectra, we find that all 6 \oiii\ emitters and 16 out of the 18 \oii\ emitters show the corresponding emission lines within the wavelength range captured by the F658N filter. The remaining two \oii\ emitters are confirmed by the relatively shallow MUSE-WIDE survey, which have their strong \oiii\ lines. 
For the other 14 sources at $z<1$, we analyze both their spectra and their broadband and narrowband images. 
Our analysis reveals no significant emission lines within the F658N filter’s wavelength range in their spectra. 
Further examination of their broadband and narrowband images shows that the majority of these sources (10/14) appear compact in narrowband but extended in broadband. 
Since the narrowband image serves as the detection image (see Section~\ref{subsec:photometry}), this discrepancy results in an underestimation of broadband flux, leading to an overestimation of the broadband-over-narrowband color, thereby erroneously categorizing them as ELG candidates. 
The remaining 4 objects appear compact in both broadband and narrowband images. Given their redshifts, the Balmer decrement and the 4000 \AA\ break contribute to a reduced flux in F606W, enhancing the broadband-over-narrowband color and consequently, misidentifying these objects as ELG candidates.

We have cross-matched 23 ELG candidates with spectroscopic redshifts in the range of $1 \lesssim z<4$. 
Upon inspecting their spectra, we identify emission lines within the wavelength range of the F658N filter in 4 objects.
These include 3 Mg\,{\sc ii}] emitter at $z\sim 1.3$ and one C\,{\sc iii}] emitter at $z\sim 2.4$.
Additionally, we also confirm one C\,{\sc iv} emitter at $z\sim 3.19$ based on the broad line partially covered by the F658N filter.
This object is matched with a QSO, CXOCDFS J033242.8-274702 from \citet{Luo2017}.
For the other 18 candidates, we thoroughly examine their spectra, broadband, and narrowband images.
5 of these candidates at $z \sim 1$ show a red continuum, inadvertently enhancing their broadband-over-narrowband colors and leading to their inaccurate selection as ELG candidates.
Among the remaining 13, we find no continuum in their spectra.
A closer look at their images reveals 8 of them to be compact in narrowband yet extended in broadband,  which would overestimate the broadband-over-narrowband color, causing them to be mistakenly identified as ELG candidates.
The final 5 objects, compact in both imaging modes, likely have large F606W-F658N colors due to continuum features that their spectra are too shallow to reveal.

It is also noticeable that faint ELG candidates from \hdha, especially for those fainter than 25 mag in F658N, would contaminate the selection of the ELG candidates. For example, when improving the significance cut of the color excesses in the ELG selection from 3 to 5, we exclude 14 ELG candidates with redshifts, which are all contamination and fainter than 25 mag in the F658N band (see panel (c) of Fig.~\ref{fig:the_redshift_color}).

\begin{figure*}[h]
    \centering
    \includegraphics[width=1\textwidth]{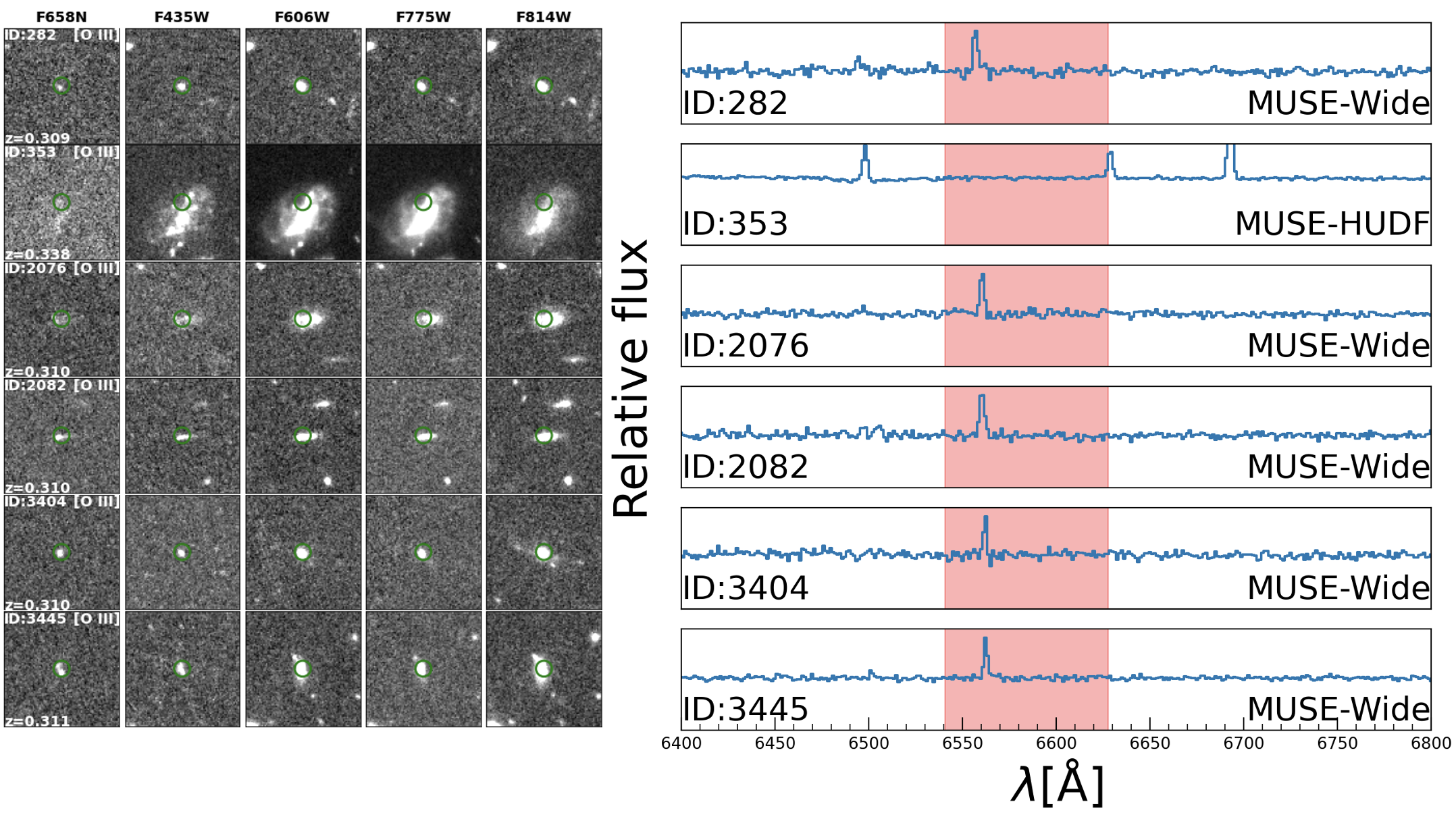}       
    \caption{
    Left panel: The narrowband and broadband thumbnail images of \oiii\ emitters at $z\sim$ 0.3. 
    Each thumbnail image has a size of $~\sim 5^{\prime\prime} \times 5^{\prime \prime}$, and a green circle with a diameter of $0^{\prime \prime}.7$ in the center. The band information (from left to right: F658N, F435W, F606W, F775W, F814W) of each column is labeled at the top. The emission line type, the catalog ID, and the redshift value are labeled in the first column for each source. 
    Right panel:
    The 1D spectra of \oiii\ emitters at $z\sim 0.3$. The shaded regions demonstrate the coverage of the F658N filter. The facilities used to obtain the spectra are marked.}
    \label{fig:oiii}
\end{figure*}

\begin{figure*}[h]
    \centering
    \includegraphics[width=0.9\textwidth]{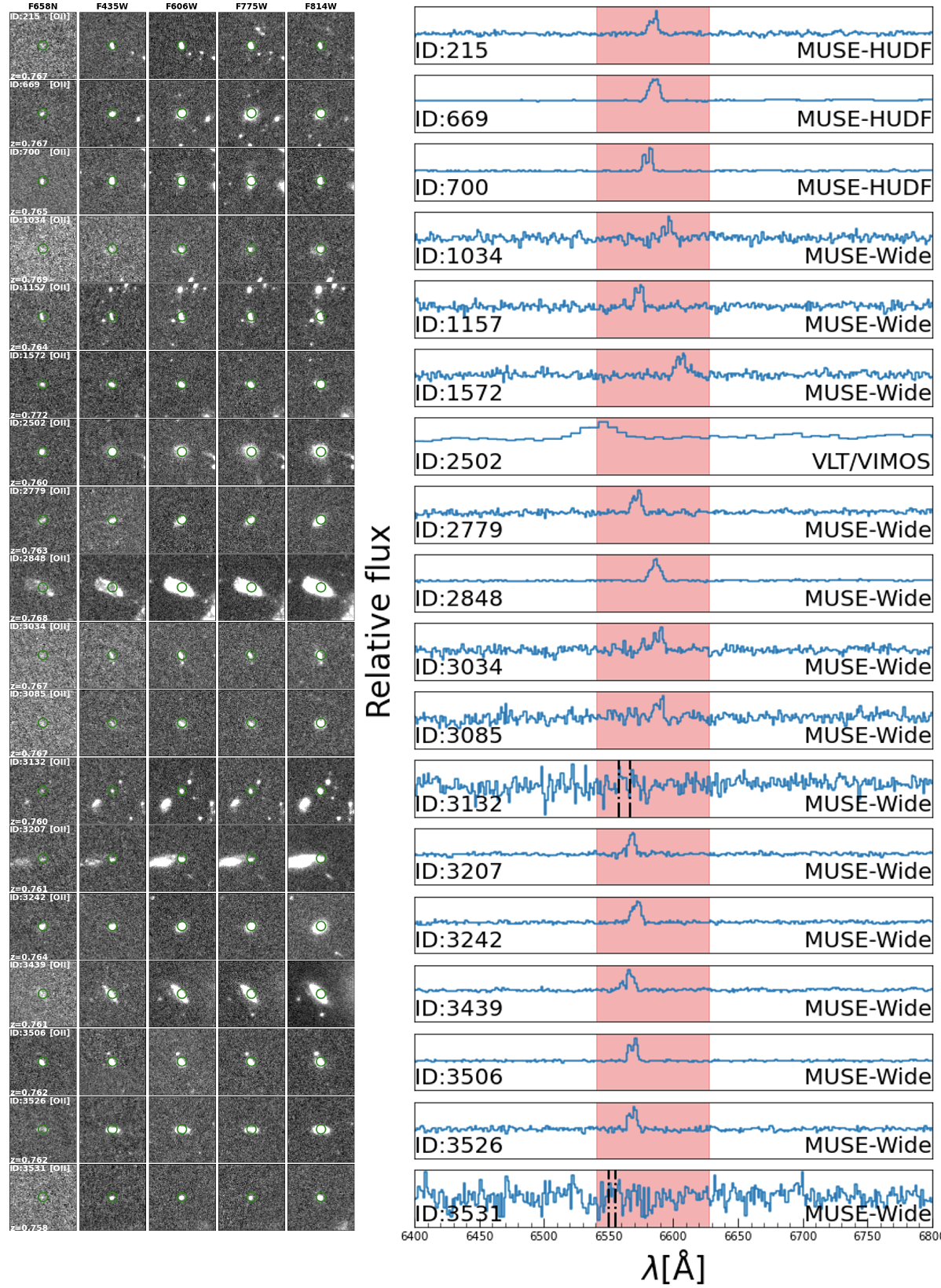}       
    \caption{Same as Fig.~\ref{fig:oiii} but for \oii\ emitters at $z\sim$ 0.75-0.77. 2 out of 18 objects (ID: 3132, ID: 3531) do not show \oii\ lines in their spectra but are confirmed by their \oiii\ lines. Their \oii\ lines in the wavelength locations are marked by the dashed black lines.
    }
    \label{fig:oii-2}
\end{figure*}

In summary, of all the ELG candidates at $z<$ 4, there are 6 and 18 galaxies confirmed as \oiii\ emitters at $z\sim$ 0.32 and \oii\ emitters at $z\sim$ 0.77, respectively. 
Their positions and photometric data are presented in Tab.~\ref{tab:elg-cat}. 
We also present these sources' narrowband and broadband thumbnail images and 1-d spectra in Fig.~\ref{fig:oiii} and Fig~\ref{fig:oii-2}.

\begin{deluxetable*}{cccccccccccc}
\tabletypesize{\scriptsize}
\label{tab:elg-cat}
\tablecaption{The catalog of \oiii\ and \oii\ emission line galaxies with spectroscopic confirmation in the \hdha\ project.}
\tablewidth{\linewidth}
\tablecolumns{10}
\tablehead{\colhead{ID} & \colhead{R.A.} & \colhead{Dec.} & \colhead{F435W} & \colhead{F606W} & \colhead{F658N} &\colhead{F775W} & \colhead{F814W} & \colhead{$z$} & \colhead{Reference} & \colhead{Line flux} &\colhead{Type}\\
\colhead{} & \colhead{(Degree)} & \colhead{(Degree)} & \colhead{(ABmag)} & \colhead{(ABmag)} & \colhead{(ABmag)} & \colhead{(ABmag)} & \colhead{(ABmag)} & \colhead{} & \colhead{} & \colhead{($10^{-17} \mathrm{erg\ cm^2\ s^{-1}}$}) &\colhead{}}
\startdata
$282^{*}$ & 53.114915 & -27.765746 & $25.93 \pm 0.03$ & $25.29 \pm 0.02$ & $24.43 \pm 0.12$ & $25.20 \pm 0.03$ & $25.07 \pm 0.01$ & $0.309$ & U19 & $1.70^{+0.47}_{-0.42}$ & \oiii\\
353 & 53.152608 & -27.769387 & $26.93 \pm 0.02$ & $25.97 \pm 0.00$ & $25.06 \pm 0.14$ & $25.63 \pm 0.00$ & $25.53 \pm 0.02$ & $0.338$ & M05 & $1.00^{+0.29}_{-0.26}$ & \oiii \\
2076 & 53.096750 & -27.819529 & $25.97 \pm 0.05$ & $25.34 \pm 0.01$ & $24.41 \pm 0.06$ & $25.22 \pm 0.03$ & $25.22 \pm 0.01$ & $0.310$ & U19 & $1.82^{+0.24}_{-0.23}$ & \oiii \\
2082 & 53.093190 & -27.819698 & $26.05 \pm 0.05$ & $25.33 \pm 0.01$ & $24.37 \pm 0.05$ & $25.17 \pm 0.03$ & $25.13 \pm 0.01$ & $0.310$ & U19 & $1.98^{+0.23}_{-0.22}$ & \oiii \\
3404 & 53.083948 & -27.858922 & $25.83 \pm 0.06$ & $24.98 \pm 0.01$ & $24.20 \pm 0.06$ & $24.82 \pm 0.03$ & $24.70 \pm 0.01$ & $0.310$ & U19 & $1.86^{+0.30}_{-0.28}$ & \oiii \\
3445 & 53.084732 & -27.861410 & $25.33 \pm 0.04$ & $24.57 \pm 0.01$ & $23.63 \pm 0.03$ & $24.45 \pm 0.02$ & $24.33 \pm 0.00$ & $0.311$ & U19 & $3.86^{+0.29}_{-0.28}$ & \oiii \\
215 & 53.167378 & -27.762830 & $26.84 \pm 0.02$ & $26.51 \pm 0.01$ & $25.33 \pm 0.16$ & $26.04 \pm 0.01$ & $25.87 \pm 0.03$ & $0.767$ & B22 & $0.97^{+0.27}_{-0.24}$ & \oii \\
669 & 53.170724 & -27.782031 & $26.38 \pm 0.01$ & $25.30 \pm 0.00$ & $24.31 \pm 0.06$ & $24.28 \pm 0.00$ & $24.12 \pm 0.01$ & $0.767$ & B22 & $2.15^{+0.25}_{-0.23}$ & \oii \\
700 & 53.152808 & -27.782777 & $25.53 \pm 0.01$ & $25.27 \pm 0.00$ & $24.18 \pm 0.04$ & $24.89 \pm 0.00$ & $24.68 \pm 0.01$ & $0.765$ & V08 & $2.63^{+0.20}_{-0.19}$ & \oii \\
1034 & 53.107234 & -27.792313 & $26.33 \pm 0.07$ & $26.02 \pm 0.03$ & $25.14 \pm 0.13$ & $25.54 \pm 0.05$ & $25.37 \pm 0.02$ & $0.769$ & U19 & $0.89^{+0.28}_{-0.25}$ & \oii \\
1157 & 53.097031 & -27.795813 & $25.71 \pm 0.04$ & $25.12 \pm 0.01$ & $24.27 \pm 0.05$ & $24.38 \pm 0.01$ & $24.26 \pm 0.01$ & $0.764$ & V08 & $1.91^{+0.23}_{-0.22}$ & \oii \\
1572 & 53.113281 & -27.807036 & $25.43 \pm 0.03$ & $25.13 \pm 0.01$ & $24.32 \pm 0.05$ & $24.52 \pm 0.02$ & $24.41 \pm 0.01$ & $0.772$ & U19 & $1.74^{+0.24}_{-0.23}$ & \oii \\
2502 & 53.030231 & -27.829806 & $24.07 \pm 0.00$ & $23.71 \pm 0.00$ & $22.92 \pm 0.02$ & $23.16 \pm 0.00$ & $23.08 \pm 0.00$ & $0.760$ & LF05 & $6.11^{+0.33}_{-0.32}$ & \oii \\
2779 & 53.073088 & -27.837126 & $25.30 \pm 0.04$ & $25.08 \pm 0.01$ & $24.27 \pm 0.05$ & $24.42 \pm 0.02$ & $24.35 \pm 0.00$ & $0.763$ & U19 & $1.82^{+0.22}_{-0.21}$ & \oii \\
2848 & 53.177903 & -27.839265 & $25.44 \pm 0.02$ & $24.53 \pm 0.01$ & $23.78 \pm 0.04$ & $23.58 \pm 0.01$ & $23.47 \pm 0.00$ & $0.768$ & R07 & $2.63^{+0.32}_{-0.31}$ & \oii \\
3034 & 53.168674 & -27.844082 & $26.40 \pm 0.06$ & $26.08 \pm 0.03$ & $25.01 \pm 0.13$ & $25.47 \pm 0.04$ & $25.29 \pm 0.01$ & $0.767$ & U19 & $1.20^{+0.30}_{-0.27}$ & \oii \\
3085 & 53.163797 & -27.845878 & $26.26 \pm 0.05$ & $26.05 \pm 0.03$ & $24.90 \pm 0.10$ & $25.42 \pm 0.04$ & $25.24 \pm 0.01$ & $0.767$ & U19 & $1.42^{+0.27}_{-0.25}$ & \oii \\
3132 & 53.091967 & -27.847772 & $26.63 \pm 0.09$ & $26.11 \pm 0.03$ & $25.02 \pm 0.11$ & $25.49 \pm 0.04$ & $25.28 \pm 0.01$ & $0.760$ & U19 & $1.21^{+0.27}_{-0.24}$ & \oii \\
3207 & 53.087621 & -27.850664 & $25.82 \pm 0.04$ & $25.58 \pm 0.02$ & $24.52 \pm 0.07$ & $25.21 \pm 0.04$ & $25.07 \pm 0.01$ & $0.761$ & U19 & $1.89^{+0.26}_{-0.24}$ & \oii \\
3242 & 53.088701 & -27.851943 & $24.51 \pm 0.01$ & $23.93 \pm 0.00$ & $23.16 \pm 0.02$ & $23.24 \pm 0.01$ & $23.13 \pm 0.00$ & $0.764$ & P09;B10 & $4.78^{+0.26}_{-0.26}$ & \oii \\
3439 & 53.126783 & -27.861306 & $25.02 \pm 0.02$ & $24.19 \pm 0.00$ & $23.37 \pm 0.04$ & $23.22 \pm 0.01$ & $23.09 \pm 0.00$ & $0.761$ & P09;B10 & $4.23^{+0.40}_{-0.39}$ & \oii \\
3506 & 53.113364 & -27.864067 & $24.25 \pm 0.01$ & $24.04 \pm 0.00$ & $22.92 \pm 0.02$ & $23.77 \pm 0.01$ & $23.59 \pm 0.00$ & $0.762$ & U19 & $8.53^{+0.31}_{-0.30}$ & \oii \\
3526 & 53.076115 & -27.865446 & $25.09 \pm 0.01$ & $24.73 \pm 0.01$ & $23.86 \pm 0.05$ & $24.18 \pm 0.01$ & $24.07 \pm 0.00$ & $0.762$ & U19 & $2.85^{+0.33}_{-0.32}$ & \oii \\
3531 & 53.087015 & -27.865916 & $26.53 \pm 0.04$ & $26.11 \pm 0.03$ & $25.15 \pm 0.16$ & $25.46 \pm 0.05$ & $25.40 \pm 0.01$ & $0.758$ & U19 & $0.96^{+0.35}_{-0.30}$ & \oii \\
\enddata
\tablenotetext{$*$}{This source has an X-Ray counterpart, and is classified as an AGN (CXOCDFS J033227.5-274555) by \cite{Luo2017}.}
\tablenotetext{ }{Redshift Reference: [1]LF05 = \citet{Le_F2005}, [2]M05= \citet{Mignoli2005}, [3]R07 = \citet{Ravikumar2007}, [4]V08 = \citet{Vanzella2008}, [5]P09;B10 = \citet{Popesso2009, Balestra2010}, [6]U19 = \citet{Urrutia2019}, [7]B22 = \citet{Bacon2022}.}
\end{deluxetable*}

\subsubsection{Confirmed LAEs at $z\simeq 4.4$}
\label{sec:zelg:comlae}

There are 16 out of 53 LAE candidates ($30\%$) matched with the spectroscopic catalogs introduced in Section~\ref{subsec:speccat}. Among those 16 sources, 13 LAE candidates are finally confirmed as LAEs at $z\sim 4.4$ with the ground-based spectroscopic surveys (see Tab.~\ref{tab:lae-cat}).
Note that one LAE confirmed by the MUSE-HUDF survey is classified as an \oiii\ emitter by the MUSE-Wide survey mistakenly.
There are also three sources at $z \simeq 4.0$ selected as LAE candidates. From their spectra, these three sources may have weak [O\,{\sc i}] lines covered by the F658N filter. We therefore exclude these 3 candidates as true LAEs.
The final success rate of the LAE candidates' confirmation is 81.25\%, which is slightly higher than the typical value of $\sim 70\%$ \citep[e.g.,][]{zheng2013,Hu2017,Harish2022} in previous works based on the ground-based telescope.

Compared to the sample of ELG at $z<4$ candidates, the sample of LAE candidates has a lower fraction in the spectroscopic coverage but a higher fraction in the spectroscopic confirmation.
The low fraction of the spectroscopic coverage is caused by the relatively faint continuum of these LAE candidates. Most of such sources have narrowband magnitudes in the range of 25$-$26, and their colors are very large (see Fig.~\ref{fig:lf-color}). Thus, the continuum of these sources is often faint, which prevents the spectroscopic follow-up surveys that often target bright broadband sources.

\begin{figure*}[h]
    \centering
    \includegraphics[width=\textwidth]{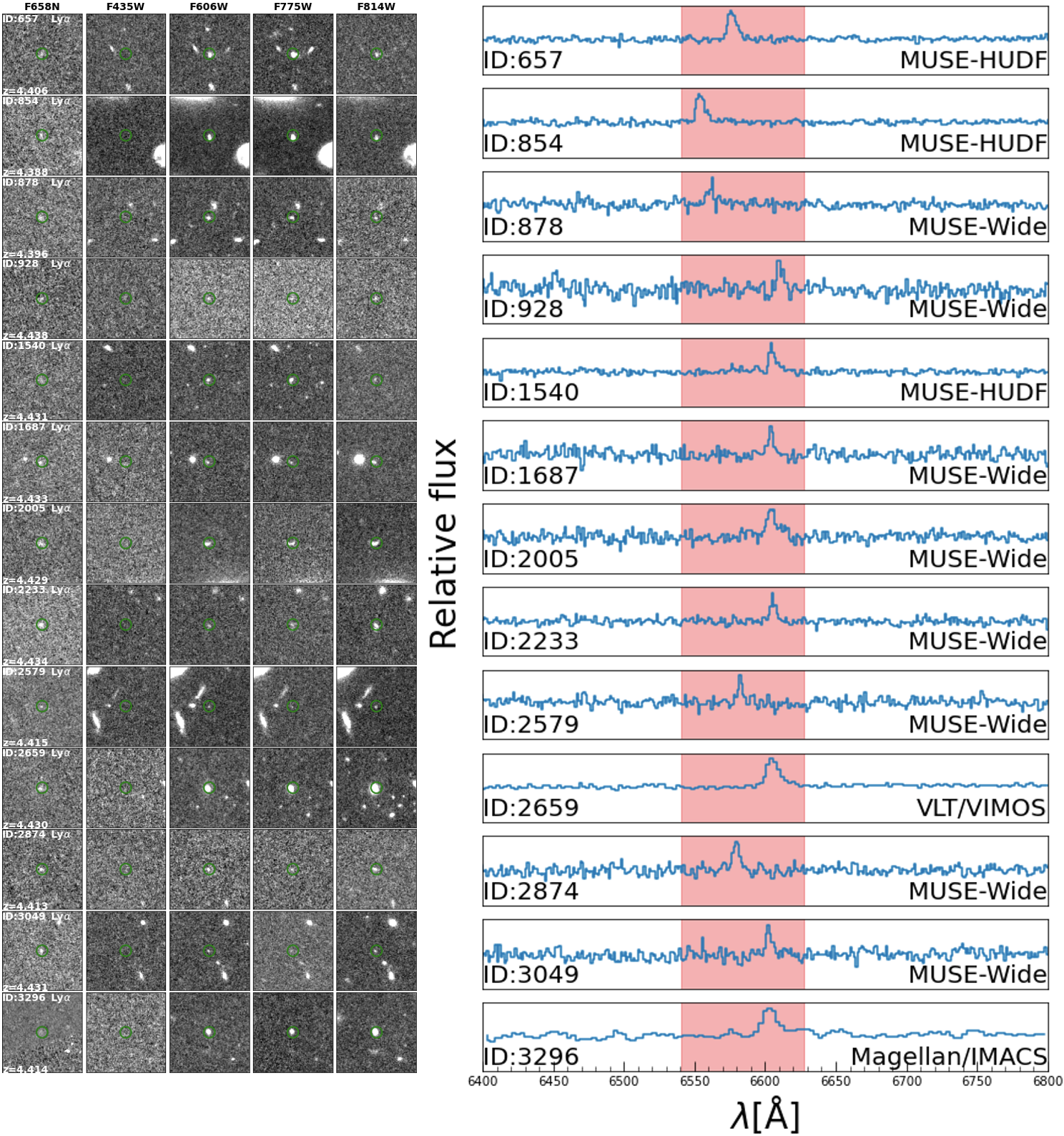}       
    \caption{Same as Fig.~\ref{fig:oiii} but for \lya\ emitters at $z\sim$ 4.4.}
    \label{fig:lae-1}
\end{figure*}

Finally, there are 13 LAE candidates confirmed as LAEs at $z\sim 4.4$. 
These sources' positions and photometric data are presented in Tab.~\ref{tab:lae-cat}.
The narrowband and broadband thumbnail images and 1-d spectra of LAEs are shown in Fig.~\ref{fig:lae-1}.

\begin{deluxetable*}{cccccccccccc}
\tabletypesize{\scriptsize}
\label{tab:lae-cat}
\tablecaption{The catalog of \lya\ emitters with spectroscopic confirmation  in the \hdha\ project.}
\tablewidth{\linewidth}
\tablecolumns{10}
\tablehead{\colhead{ID} & \colhead{R.A.} & \colhead{Dec.} & \colhead{F435W} & \colhead{F606W} & \colhead{F658N} &\colhead{F775W} & \colhead{F814W} & \colhead{$z$} & \colhead{Reference} & \colhead{Line flux} &\colhead{Type}\\
\colhead{} & \colhead{(Degree)} & \colhead{(Degree)} & \colhead{(ABmag)} & \colhead{(ABmag)} & \colhead{(ABmag)} & \colhead{(ABmag)} & \colhead{(ABmag)} & \colhead{} & \colhead{} & \colhead{($10^{-17} \mathrm{erg\ cm^2\ s^{-1}}$}) &\colhead{}}
\startdata
657 & 53.181079 & -27.781187 & $-99.00^{*}$ & $28.09 \pm 0.03$ & $25.41 \pm 0.14$ & $27.30 \pm 0.02$ & $27.03 \pm 0.09$ & $4.411$ & B22 & $1.39^{+0.21}_{-0.19}$ & Ly$\alpha$ \\
854 & 53.192930 & -27.786922 & $-99.00^{*}$ & $27.86 \pm 0.02$ & $25.85 \pm 0.19$ & $26.95 \pm 0.01$ & $27.07 \pm 0.09$ & $4.393$ & B22 & $0.84^{+0.21}_{-0.17}$ & Ly$\alpha$ \\
878 & 53.137240 & -27.787397 & $34.11 \pm 15.91$ & $28.57 \pm 0.04$ & $25.60 \pm 0.16$ & $28.76 \pm 0.07$ & $29.29 \pm 0.56$ & $4.399$ & B22 & $1.20^{+0.21}_{-0.18}$ & Ly$\alpha$ \\
928 & 53.114967 & -27.789050 & $28.80 \pm 0.70$ & $28.91 \pm 0.62$ & $25.68 \pm 0.20$ & $27.93 \pm 0.41$ & $28.52 \pm 0.31$ & $4.438$ & U19 & $1.13^{+0.28}_{-0.26}$ & Ly$\alpha$ \\
1540 & 53.175746 & -27.806146 & $-99.00^{*}$ & $28.84 \pm 0.05$ & $25.89 \pm 0.19$ & $28.20 \pm 0.04$ & $27.94 \pm 0.18$ & $4.434$ & B22 & $0.92^{+0.20}_{-0.17}$ & Ly$\alpha$ \\
1687 & 53.110863 & -27.809801 & $28.04 \pm 0.33$ & $27.29 \pm 0.10$ & $25.20 \pm 0.11$ & $26.97 \pm 0.16$ & $27.06 \pm 0.08$ & $4.433$ & U19 & $1.54^{+0.24}_{-0.22}$ & Ly$\alpha$ \\
$2005^{\dagger}$ & 53.204212 & -27.817288 & $29.44 \pm 0.67$ & $26.95 \pm 0.06$ & $24.76 \pm 0.08$ & $26.45 \pm 0.10$ & $26.25 \pm 0.04$ & $4.429$ & F11;G21 & $2.36^{+0.23}_{-0.22}$ & Ly$\alpha$ \\
2233 & 53.128212 & -27.823525 & $-99.00^{*}$ & $27.45 \pm 0.17$ & $25.28 \pm 0.11$ & $26.64 \pm 0.12$ & $26.65 \pm 0.05$ & $4.434$ & U19 & $1.46^{+0.22}_{-0.21}$ & Ly$\alpha$ \\
2579 & 53.175339 & -27.831462 & $30.00 \pm 1.53$ & $28.58 \pm 0.29$ & $25.56 \pm 0.21$ & $28.18 \pm 0.53$ & $28.17 \pm 0.26$ & $4.415$ & U19 & $1.25^{+0.31}_{-0.27}$ & Ly$\alpha$ \\
$2659^{\dagger}$ & 53.225164 & -27.833627 & $-99.00^{*}$ & $25.78 \pm 0.01$ & $24.28 \pm 0.08$ & $24.87 \pm 0.01$ & $24.84 \pm 0.00$ & $4.430$ & F11;P09;B10 & $3.02^{+0.34}_{-0.32}$ & Ly$\alpha$ \\
2874 & 53.162809 & -27.839717 & $28.78 \pm 0.50$ & $27.87 \pm 0.14$ & $25.37 \pm 0.14$ & $27.23 \pm 0.22$ & $27.49 \pm 0.10$ & $4.413$ & U19 & $1.42^{+0.24}_{-0.22}$ & Ly$\alpha$ \\
3049 & 53.159479 & -27.844401 & $29.10 \pm 0.68$ & $29.27 \pm 0.52$ & $25.83 \pm 0.22$ & $27.83 \pm 0.38$ & $28.65 \pm 0.23$ & $4.431$ & U19 & $1.00^{+0.26}_{-0.23}$ & Ly$\alpha$ \\
$3296^{\dagger}$ & 53.165752 & -27.854247 & $29.49 \pm 0.97$ & $26.36 \pm 0.03$ & $24.37 \pm 0.11$ & $25.54 \pm 0.05$ & $25.42 \pm 0.01$ & $4.414$ & F11 & $3.25^{+0.47}_{-0.43}$ & Ly$\alpha$ \\
\enddata
\tablenotetext{$*$}{Here -99 means non-detection in the F435W band.}
\tablenotetext{\dagger}{LAEs are also reported by \citet{Finkelstein2011}.}
\tablenotetext{ }{Redshift Reference: [1]P09;B10 = \citet{Popesso2009,Balestra2010}, [2]F11 = \citet{Finkelstein2011}, [3]U19 = \citet{Urrutia2019}, [4]G21 = \citet{Garilli2021}, [5]B22 = \citet{Bacon2022}.}
\end{deluxetable*}
    
\subsection{Depths of Emission Lines in \hdha}
\label{sec:linedepth}

\begin{figure*}[ht]
    \centering
    \includegraphics[width=\textwidth]{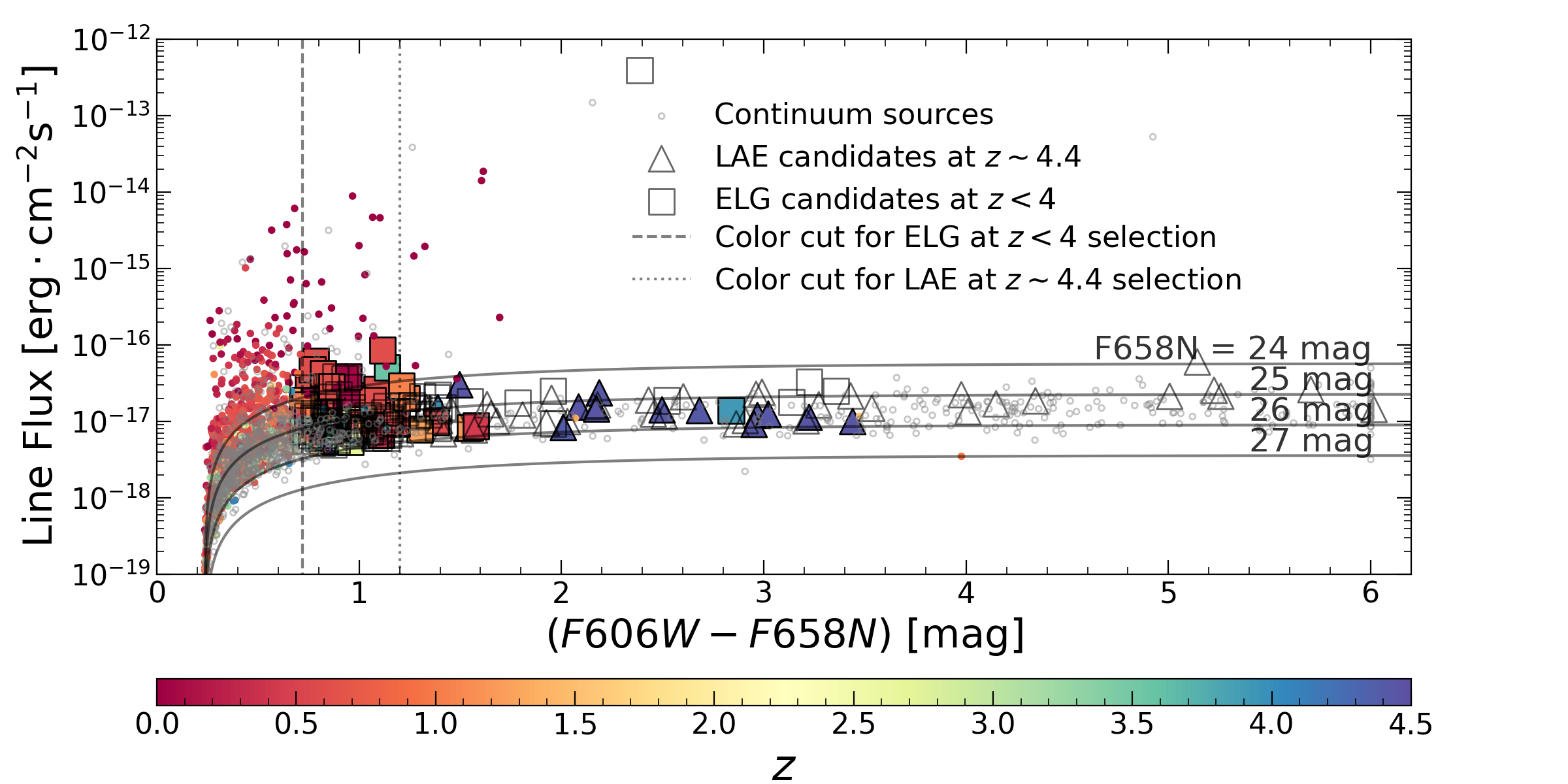}
    \caption{The line flux and F606W-F658N color diagram of the continuum-dominated sources, 
    ELG candidates at $z<4$, and LAE candidates at $\sim 4.4$ in the \hdha.
    The symbols are similar to those in Fig. \ref{fig:color_sig_cmd}.
    The open symbols and color-filled symbols represent sources without and with spectroscopic redshifts, respectively.
    Their redshifts are encoded by the color bar at the bottom.
    The solid lines are the curves of NB magnitude that step from 24 to 27 mag. 
    The vertical dotted and dashed lines indicate the color cut values for ELGs at $z<4$ and LAEs at $z\sim $4.4, respectively.}
    \label{fig:lf-color}
\end{figure*}

Here we calculate the depths of emission lines of ELGs candidate in the \hdha. The emission line fluxes $f_{line}$ of \hdha\ sources are defined as:
\begin{equation}    \label{eq:lf}
     {f_{line} = \Delta \lambda_{NB} \left( \frac{f_{\lambda, NB} - f_{\lambda, BB}}{1 - \frac{\Delta \lambda_{NB}}{\Delta \lambda_{BB}}} \right)} ,
\end{equation}
where the $f_{\lambda, BB}$ and $f_{\lambda, NB}$ are the flux densities in the unit of $\mathrm{erg\ cm^2\ s^{-1}\ \AA^{-1}}$ of the F606W band and the F658N band,  and the
$\Delta \lambda_{BB}$ and $\Delta \lambda_{NB}$ are the filter widths of the broadband
and narrowband, respectively. We present the relation between the line fluxes and the colors of all sources, ELG candidates at $z < 4$, and LAE candidates at $z \sim 4.4$ in Fig.~\ref{fig:lf-color}. 
The redshifts of sources matched successfully with spectroscopic catalogs are encoded by the color bar. We also plot the different limiting magnitudes in the F658N band in Fig.~\ref{fig:lf-color}, which shows that the minimum line fluxes of the candidates depend on both the color cut and the limiting F658N magnitude we used in selecting ELGs at $z<4$ or LAEs $z\sim 4.4$.
With a smaller color cut and deeper F658N data, a fainter line flux can be probed. 
We define the depths of our ELG candidates at $z<4$ (LAE candidates at $z\sim 4.4$) as the minimum line fluxes when the color cut is set to 0.72 (1.2) and the limiting F658N magnitude is set to 25.8. Then the depths of emission lines in the \hdha\ project are $3.9 \times 10 ^{-18} \mathrm{erg\ cm^2\ s^{-1}}$ and $6.4 \times 10^{-18} 
\mathrm{erg\ cm^2\ s^{-1}}$ for ELG candidates at $z< 4$ and LAE candidates at $z\sim 4.4$, respectively. For those confirmed ELGs (LAEs), the minimum and median line fluxes are $8.9 \times 10^{-18} \mathrm{erg\ cm^2\ s^{-1}}$ and $1.9 \times 10^{-17} \mathrm{erg\ cm^2\ s^{-1}}$ ($8.4 \times 10^{-18} \mathrm{erg\ cm^2\ s^{-1}}$ and $1.4 \times 10^{-17} \mathrm{erg\ cm^2\ s^{-1}}$), respectively. 

\section{SUMMARY}
\label{sec:sum}
We present the Hubble Deep Hydrogen Alpha (HDH$\alpha$) project, which introduces the deepest narrowband imaging data with \textit{HST}. The \hdha\ field covers a partial area of the GOODS-S field with the \textit{HST}/ACS F658N imaging. This paper is the first \hdha\ paper in a series.

We have reduced the F658N narrowband imaging data in the GOODS-S field. All images of the \hdha\ are checked, aligned, and drizzled into the same pixel grid of the HLF. To ensure the \hdha\ data is reliable enough for the photometric analysis, we carry out tests on the PSF and the astrometry of our image.

With the \hdha, we select 156 ELG candidates at $z<4$ and 53 LAE candidates at $z\sim4.4$. We further cross match these two samples with the public spectroscopic catalogs and find 61 ELG candidates at $z<4$ and 16 LAE candidates at $z \sim 4.4$  matched. After validating their redshifts, 6 \oiii\ emitters, 18 \oii\ emitters, and 13 LAEs are confirmed, which are listed in Tab.~\ref{tab:elg-cat} and \ref{tab:lae-cat}.

The \hdha\ image increases the legacy value of the GOODS-S field by adding the first narrowband image to the existing data sets. The narrowband images based on \hst\ enable to search for LAEs with a very faint continuum, whose continuum flux may be out of reach by ground-based broadband images. It is also the best tool to understand the morphology of emission line regions of galaxies and to statistically probe the line properties, such as the line luminosity functions and equivalent width distributions, as well as their morphology.     

\vspace{15mm}

\section*{acknowledgments}
We would like to thank the referee for very helpful comments and suggestions, which significantly improved this paper. Z.Y.Z. acknowledges the support of the National Key R\&D Program of China No.2022YFF0503402, the National Science Foundation of China (12022303), and the China-Chile Joint Research Fund (CCJRF No. 1906). J.X.W. thanks the support from the National Science Foundation of China (11890693). We acknowledge the science research grants
from the China Manned Space Project with No. CMS-CSST-2021-A07, CMS-CSST-2021-A04 and CMS-CSST-2021-B04.

This work is based on observations taken by the 3D-HST Treasury Program (GO 12177 and 12328) with the NASA/ESA HST, which is operated by the Association of Universities for Research in Astronomy, Inc., under NASA contract NAS5-26555.
This work is based on observations taken by the MUSE-Wide Survey as part of the MUSE Consortium.
Based on observations made at the European Southern Observatory, Paranal, Chile (ESO LP 164.O-0560).
Based on observations made with ESO Telescopes at the La Silla or Paranal Observatories under programme ID 171.A-3045.
Based on data obtained with the European Southern Observatory
Very Large Telescope, Paranal, Chile, program 070.A-9007(A)
Observations have been carried out using the Very Large Telescope at the ESO Paranal Observatory under Program ID(s): 170.A-0788, 074.A-0709, and 275.A-5060
Based on data products created from observations collected at the European Organisation for Astronomical Research in the Southern Hemisphere under ESO programme 194.A-2003(E-T).

\vspace{5mm}

\facilities{\hst\ (ACS)}
\software{Astropy \citep{astropy:2013,astropy:2018}, 
                    Drizzlepac \citep{Gonzaga2012}, 
                    HEALPix and Healpy \citep{2005ApJ...622..759G,Zonca2019},
                    Matplotlib \citep{Hunter:2007}, 
                    Numpy \citep{harris2020array}, 
                    Photutils \citep{larry_bradley_2021_4624996}, 
                    Regions.
                    Source Extractor \citep{Bertin1996},
                    SWarp \citep{Bertin2010},
                    Scipy \citep{2020SciPy-NMeth}, 
                    }
\clearpage

\appendix
\label{sec:appendix}

\section*{Star contamination in the \ha\ emitter candidates in \hdha}
To exclude stars in the sample of ELG candidates in \hdha, we follow the star selection method introduced in \citep{Dahlen2010}, which is based on the F435W-J color and the J-IRAC1 color. 
Here the F435W band magnitudes for all 208 ELG candidates are measured using the method described in Section \ref{sec:photo}. 
For the J band and IRAC1 band magnitudes, we use the measurements from the 3D-HST project\citep{Skelton14}.

There are 186 objects successfully matched with the 3D-HST catalog with the J band and IRAC1 band magnitudes. 
For the remaining 22 objects, we measure the magnitudes in the J band image and the IRAC1 image directly from the TENIS program \citep{Hsieh2012} and the SEDs program \citep{Ashby2013}, respectively. 
We use apertures with a radius of 2$^{\prime \prime}$ in the J band and IRAC1 band measurements.
We also correct the magnitude zero-point offset for different programs in the J band and the IRAC1 band for consistent color measurements.
There are 178 objects detected at $>2\sigma$ level in both the J band image and IRAC1 image.
6 and 11 objects are detected at only the J band and the IRAC1 band, respectively. We use the 2$\sigma$ up limit magnitude for the non-detection in the corresponding band. The remaining 13 objects have no detection in neither J band nor IRAC1 band images. However, most of their HST images show faint and extended features, thus these 13 objects are not classified as stars.

\begin{figure}
    \centering
    \includegraphics[width=0.7 \textwidth]{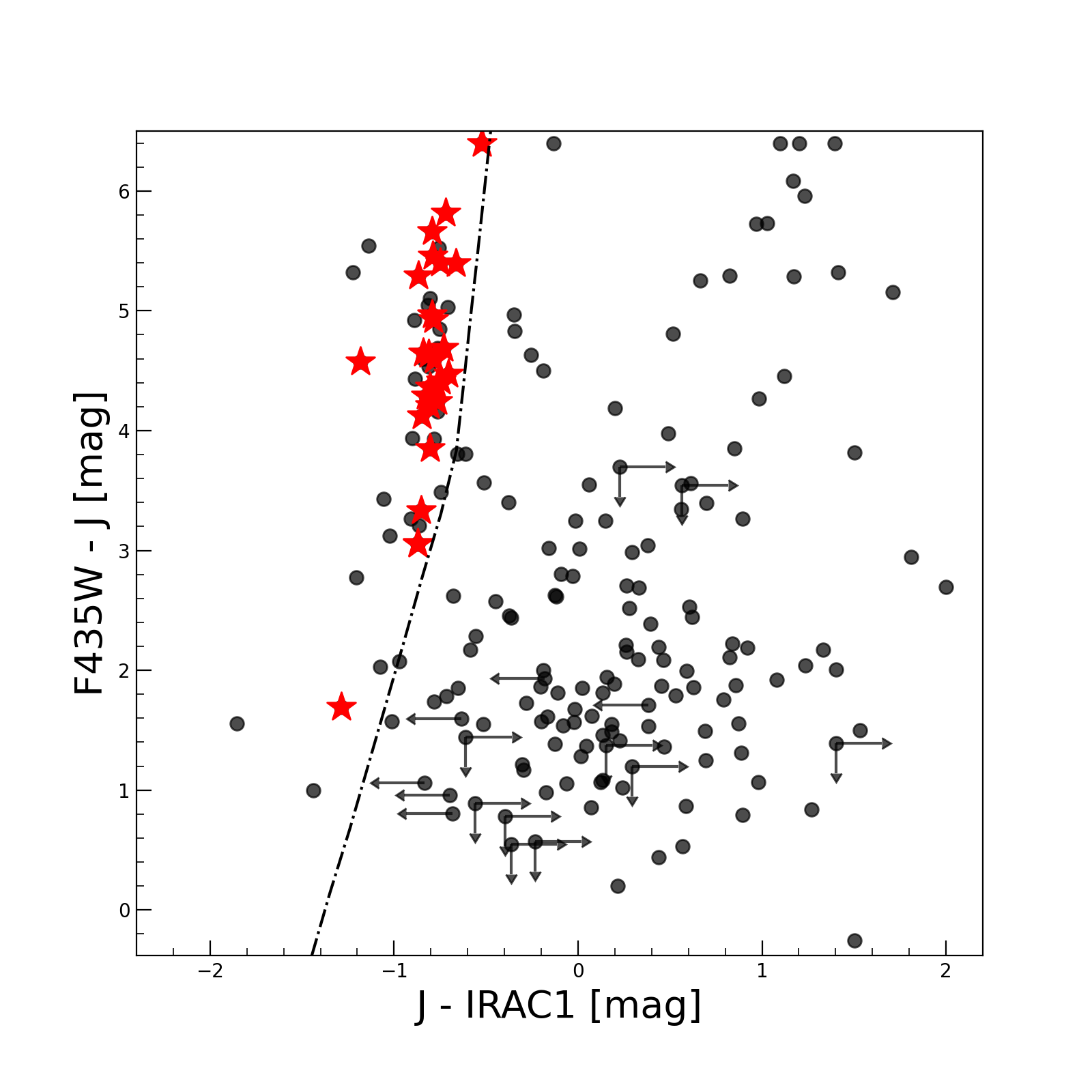}
    \caption{The color-color diagram used for selecting stars from our \hdha\ ELG candidates. The dash line represents the star selection criteria used by \citet{Dahlen2010}.
    ELG candidates with S/N $>2$ in the J band and IRAC1 band are plotted as filled circles in this figure, and ELG candidates spectroscopically confirmed as stars are highlighted as red asterisks. }
    \label{fig:star-selection}
\end{figure}

We present the color-color diagram of ELGs in Fig.~\ref{fig:star-selection}.
The dashed line presents our selection criteria of stars, which is based on the selection criteria of \citet{Dahlen2010}.
For the 17 objects detected at either J or IRAC bands (the upper or lower limit arrows in Fig.~\ref{fig:star-selection}), none are classified as stars according to the star selection method, and 6 of the 17 objects have available redshift information with $0.62<z<3.97$. 
For 178 objects detected at $>2 \sigma$ in both images (filled circles in Fig.~\ref{fig:star-selection}), 52 objects are classified as stars.
Among these 52 objects, only 27 objects have available redshift information. 
All these 27 objects are identified as stars (asterisks in Fig.~\ref{fig:star-selection}) based on their spectra, implying the success of the star selection method. Furthermore, most of the remaining 25 objects show point-source morphology in the HST images. Therefore, these 52 objects are excluded as the ELG candidates in \hdha.

\clearpage

\bibliography{reference}{}
\bibliographystyle{aasjournal}

\end{document}